\documentclass[%
prd,
endfloats*,
showpacs,
showkeys,
preprint,
nofootinbib 
] {revtex4}
\usepackage{graphicx,hyperref,amssymb,amsmath,enumerate}

\def\mue{$(\mu^-, e^+)\ $}

\begin{document}
\title{Nuclear $(\mu^-,e^+)$ conversion mediated by Majorana neutrinos}

\date{\today}

\author{P. Domin}
\email{pavol.domin@usm.cl} 
\affiliation{Departamento de F{\'i}sica, Universidad T{\' e}cnica Federico Santa Mar{\' i}a,
Casilla 110-V, Valpara{\' i}so, Chile}
\author{S. Kovalenko}
\email{sergey.kovalenko@usm.cl} 
\affiliation{Departamento de F{\'i}sica, Universidad T{\' e}cnica Federico Santa Mar{\' i}a,
Casilla 110-V, Valpara{\' i}so, Chile}
\author{Amand Faessler}
\affiliation{Institute f\"ur Theoretische Physik der Univesit\"at
T\"ubingen, Auf der Morgenstelle 14, D-72076 T\"ubingen, Germany}

\author{F. \v Simkovic}
\email{fedor.simkovic@fmph.uniba.sk}
\altaffiliation{On  leave of absence from Department of Nuclear
Physics, Comenius University, Mlynsk\'a dolina F1, SK--842 15
Bratislava, Slovakia} \affiliation{Institute f\"ur Theoretische
Physik der Univesit\"at T\"ubingen, Auf der Morgenstelle 14,
D-72076 T\"ubingen, Germany}

\begin{abstract}
  We study lepton number violating (LNV) process of $(\mu^-, e^+)$ conversion in nuclei
  mediated by the exchange of light and heavy Majorana neutrinos.
  Nuclear structure calculations have been carried out for the case of
  experimentally interesting nucleus ${^{48}\mathrm{Ti}}$ in the framework
  of renormalized proton-neutron Quasiparticle Random Phase Approximation.
  We demonstrate that the imaginary part of the amplitude of light Majorana neutrino
  exchange mechanism gives an appreciable contribution to the \mue conversion rate.
  This specific feature is absent in the allied case of $0\nu\beta\beta$ decay.
  %
    Using the present neutrino oscillations, tritium beta decay, accelerator and
  cosmological data we derived the limits on the effective masses of light
  ${\langle m \rangle_{\mu e}}$ and heavy $\langle M_{N}^{-1} \rangle_{\mu e}$ neutrinos.
  The expected rates of nuclear \mue conversion, corresponding to these limits,
  were found to be so small that even within a distant future the \mue conversion
  experiments will hardly be able to detect the neutrino signal.
  Therefore, searches for this LNV process
  can only rely on the presence of certain physics beyond the trivial extension of
  the Standard Model by inclusion of massive Majorana neutrinos.
 \end{abstract}



\maketitle

\section{\label{sec:introduction}Introduction}

Lepton number ($L$) conservation is one of the most obscure sides
of the Standard Model (SM) not supported by an underlying
principle and following from an accidental interplay between gauge
symmetry and field content. Any deviation from the SM structure
may introduce $L$ non-conservation~(LNV). Over the years the
possibility of lepton number non-conservation has been attracting
a great deal of theoretical and experimental efforts since any
positive experimental signal of LNV would point to physics beyond
the SM. The simplest extension of the SM allowing LNV processes
implies inclusion of massive Majorana neutrinos with the $\Delta
L=2$ mass term introducing the necessary source of LNV. However,
the role of neutrinos in LNV processes is more intricate. The
fundamental fact~\cite{sch82} consists in the following:
observation of any LNV process would prove that neutrinos are
massive Majorana particles. This is true even if their direct
contribution to this process is negligible and the dominant
contribution has nothing to do with neutrinos.

Recent neutrino oscillation experiments established the presence of small non-zero
neutrino masses, the fact that itself points to physics beyond the SM.  However
neutrino oscillations are not sensitive to the nature of neutrinos: they
could be either Majorana or Dirac particles leading to the same oscillation observables.

The principal question if neutrinos are Majorana or Dirac particles can be
answered only by searching for LNV processes which, as commented above, are
intimately related to the nature of neutrinos.

Various LNV processes have been discussed in the literature in this respect
(for review see~\cite{dib01}). In principle, they can probe Majorana
neutrino contribution and provide information on the so called effective masses
$\langle m_\nu \rangle_{\alpha\beta}$ and $\langle M_{N}^{-1} \rangle_{\alpha\beta}$ of
light and heavy Majorana neutrinos (for definition see Sect. \ref{sec:level2}).
These quantities under certain assumptions are related to the
entries of the Majorana neutrino mass matrix $M_{\alpha\beta}^{(\nu)}$.

Among these processes there are a few
LNV nuclear processes having prospects for experimental searches:
neutrinoless double beta decay ($0\nu\beta\beta$), muon to positron $(\mu^-,e^+)$ conversion
and, probably,  muon to antimuon $(\mu^-, \mu^+)$ conversion~\cite{mis94,sim01c}.

Currently the most sensitive experiments intended to distinguish
the Majorana nature of neutrinos are those searching for
$0\nu\beta\beta$-decay
~\cite{kla01a,aal02,arn04,ell02}. 
The nuclear theory side \cite{doi85,fae98,suh98} of this process has
been significantly improved in the last decade (see
also~\cite{ell04,rod03,bil04} and references therein) allowing
reliable extraction of fundamental particle physics parameters
from experimental data.

The $(\mu^-, e^+)$ conversion is another LNV nuclear process
searched for experimentally. The important role of muon as a test
particle for new physics beyond the SM has been recognized long
time ago.  When negative muons penetrate into matter they can be
trapped to atomic orbits. Then the bound muon may disappear either
decaying into one electron and two neutrinos or being captured by
the nucleus, i.e., due to  ordinary muon capture. These two
processes, conserving both total lepton number and lepton flavors,
are the SM processes and have been well studied both theoretically
and experimentally. The physics beyond the SM resides in yet
non-observed
channels of muon capture:  muon-electron ($\mu^-, e^-$) and
muon-positron ($\mu^-, e^+$) conversions in nuclei
\cite{kam79,ver81,ver82,leo83,kos94,kos97,fae00,sim01a,div01,div02,kos01a,kos01b,sim01b,kos02,kos03,fae04a}:
\begin{equation}
  \begin{split}
    \label{eq.1}
    (A,Z) + \mu_{\mathrm{b}}^- &\rightarrow  e^- + (A,Z)^*, \\
    (A,Z) + \mu_{\mathrm{b}}^- &\rightarrow  e^+ + (A,Z-2)^*.
  \end{split}
\end{equation}
Apparently,  the ($\mu^-, e^-$) conversion process violate lepton flavor $L_f$
and conserve the total lepton number $L$,  while ($\mu^-, e^+$) conversions
violate both of them.   Additional differences between the $(\mu^-, e^-)$ and
$(\mu^-, e^+)$ lie on the nuclear physics side. The first process can
proceed on one nucleon of the participating
nucleus while the second process involves two nucleons as dictated by charge
conservation \cite{ver81,leo83}.
Note also that the $(\mu^-, e^-)$ conversion amplitude is quadratic and $(\mu^-, e^+)$
amplitude linear in the light neutrino mass. Thus the second process looks more
sensitive to the light neutrino masses.

The currently best experimental limit on  $(\mu^-,e^+)$ conversion
branching ratio has been established at PSI ~\cite{doh93} for the
$^{48}$Ti nuclear target
\begin{align}
R^{(\mu e^+)}(Ti) =  \frac{\Gamma(\mu^-+ {^{48}\mathrm{Ti}}
  \rightarrow e^+ + {^{48}\mathrm{Ca}})}
  {\Gamma(\mu^- + {^{48}\mathrm{Ti}}
  \rightarrow \nu_{\mu} + {^{48}\mathrm{Sc}})} <
  4.3\times 10^{-12}.
\label{eq.2}
\end{align}
Now it is expected a significant improvement of this limit
in the near future experiments: SINDRUM II (PSI) with $^{48}$Ti target ~\cite{doh93},
MECO (Brookhaven) with ${}^{27}$Al target~\cite{mol02} and PRIME (Tokyo) with
$^{48}$Ti target ~\cite{kunxx}.

In the present paper we study light and heavy Majorana neutrino exchange mechanisms
of the $(\mu^-, e^+)$ conversion which are conceptually most natural and simple.
One of the main motivations of this study comes from the nuclear physics side
of this process: the nuclear theory of $(\mu^-, e^+)$ conversion is not yet
well elaborated and may show new interesting features absent in
the other LNV processes such as the $0\nu\beta\beta$-decay.
For instance, as we will demonstrate, the imaginary part of the
$(\mu^-, e^+)$ conversion amplitude in the case of light Majorana
exchange gives an appreciable contribution to the rate of this
process, the fact which has not been recognized for a long time.
Studying the most simple case of $(\mu^-, e^+)$ conversion via
Majorana neutrino exchange, we have in mind that this process may
receive contribution from other mechanisms offered by various
models beyond the SM such as the R-parity violating supersymmetric
models, the leptoquark extensions of the SM etc.  Some of these
mechanisms may involve light or heavy neutrino exchange and,
therefore, in the part of nuclear structure calculations they may
resemble the ordinary neutrino mechanisms. Thus our present study
can be viewed as a step towards a more general description of
$(\mu^-, e^+)$ conversion including all the possible mechanisms.

Below, we develop a detailed nuclear structure theory for the light and heavy
neutrino exchange mechanisms of this process on the basis of the nuclear
proton-neutron renormalized Quasiparticle Random Phase Approximation (pn-QRPA)
~\cite{toi95,sch96}.  We calculate the nuclear matrix elements of $(\mu^-, e^+)$
conversion in $^{48}$Ti, which serves as target nucleus in SINDRUM~\cite{doh93} and
PRIME~\cite{kunxx} experiments.

Existing limits on neutrino masses and mixing from neutrino
oscillation phenomenology and other observational data allow us to
estimate typical rate of this process, assuming the dominance of
light or heavy Majorana neutrino exchange mechanisms.
Extremely low values for these rates, derived in this way, leave
no chance to detect a neutrino signal in the \mue conversion even
within a distant future and, thus, to derive information on the
effective masses $\langle m_\nu \rangle_{\mu e}$ and $\langle
M_{N}^{-1} \rangle_{\mu e}$ from this process. This conclusion,
nevertheless, does not diminish the importance of experiments
searching for \mue conversion since its observation would be
unambiguous signal of a non-trivial physics beyond the SM.

The paper is organized as follows. In Sect.~\ref{sec:level2} we discuss some
general issues of Majorana neutrinos for LNV processes. Sect.~\ref{sec:level2-1}
deals with the current limits on the effective Majorana neutrino masses entering to
the $(\mu^-, e^+)$ conversion amplitude.
The amplitude and rate of $(\mu^-, e^+)$ conversion are derived in Sect.~\ref{sec:level3}. The
details of nuclear calculations for $(\mu^-,e^+)$ conversion in
$^{48}$Ti are given in Sect.~\ref{sec:level4}.
In Sect.~\ref{sec:level4-1} we discuss the possible impact of
\mue conversion experiments on neutrino physics and visa versa. In Sect.~\ref{sec:level5} we
summarize our results and conclusions.

\section{\label{sec:level2}Majorana neutrinos in LNV processes}

The finite masses of neutrinos are tightly related to the problem of  lepton
flavor/number violation. The Dirac, Majorana and Dirac-Majorana neutrino mass
terms in the Lagrangian offer different neutrino mixing schemes and allow
various lepton number/flavor violating processes \cite{bil87,bil99b,zub02}.

Let us consider the generic case of neutrino field contents with the three left-handed weak
doublet neutrinos $\nu'_{Li} = (\nu'_{Le},\nu'_{L\mu},\nu'_{L\tau})$
and $n$ species of the SM singlet right-handed neutrinos
$\nu'_{Ri}=(\nu'_{R1},...\nu'_{Rn})$.
The mass term for this set of fields can be written in a general form as
\begin{eqnarray}\label{mass-term}
\nonumber
&-& \frac{1}{2} \overline{\nu^{\prime}} {\cal M}^{(\nu)} \nu^{\prime c} +
\mbox{H.c.} =
- \frac{1}{2}
(\bar\nu'_{_L},  \overline{\nu_{_R}^{\prime c}})
\left(\begin{array}{cc}
{\cal M}_L & {\cal M}_D \\
{\cal M}^T_D  & {\cal M}_R \end{array}\right)
\left(\begin{array}{c}
\nu_{_L}^{\prime c} \\
\nu'_{_R}\end{array}\right) + \mbox{H.c.} =\\
&-&\frac{1}{2} \sum_{i=1}^{3+n} m_{i} \overline{\nu^c}_{i}\nu_i
+ \mbox{H.c.}
\end{eqnarray}
Here ${\cal M}_L, {\cal M}_R$ are $3\times 3$ and $n\times n$ symmetric
Majorana mass matrices, ${M}_D$ is $3\times n$ Dirac type matrix.
Rotating the neutrino mass matrix by the unitary transformation to the diagonal form
\begin{eqnarray}\label{rotation}
U^T {\cal M}^{(\nu)}U = Diag\{m_{i}\}
\end{eqnarray}
we end up with $n+3$ Majorana neutrinos
$\nu_i =  U^*_{ki} \nu'_{k}$ with the masses $m_{i}$.
In special cases there may
appear among them pairs with masses degenerate in
absolute values. Each of these pairs can be collected into a Dirac neutrino
field. This situation corresponds to conservation of certain lepton numbers
assigned to these Dirac fields.

The considered generic model must contain at least three observable light neutrinos
while the other states may be of arbitrary mass.  In particular, they may include
intermediate and heavy mass states. Presence or absence of these neutrino states is
a question for experimental searches.

The favored neutrino  model has to accommodate modern neutrino
phenomenology in a natural way, in particular, to answer the
question of the smallness of neutrino masses compared to the
charged lepton ones.  The most prominent guiding principle in this
problem is the see-saw mechanism. It suggests that the typical
scale of $M^D$ matrix elements in Eq. (\ref{mass-term}) is
comparable with the masses of charged leptons meanwhile  the
${\cal M}_R$ is associated to a large hypothetical scale of lepton
number violation like $M_{LNV} \approx 10^{12}~\mathrm{GeV}$. Then
the diagonalization in Eq. (\ref{rotation}) brings very light
$\nu_k$ and very heavy $N_k$ Majorana neutrinos. This mechanism
can be realized in various models beyond the SM with significantly
lower scales, $M_{LNV}\sim 1$ TeV, leading to the neutrino masses
and mixing consistent with the observational data. A particular
example is given by the class of supersymmetric model with
bilinear R-parity violation (see, for instance,
Ref.~\cite{hirsch03} and references therein). In these models the
heavy Majorana neutrinos have moderately large masses $\sim 1$TeV
and even lower giving them phenomenological significance via a
priori non-negligible contributions to LNV processes. In the
present paper we examine the contributions of light and heavy
Majorana neutrinos to $(\mu^-, e^+)$ conversion.

In general, the flavor neutrino states are the superpositions of
light ($\nu_{k}$) and heavy ($N_{k}$) Majorana mass eigenstates:
\begin{equation}
  \label{eq:numixing}
  \nu_{l}(x) = \sum_{k=\mathrm{light}} U_{lk} \nu_{k}(x) +
  \sum_{k=\mathrm{heavy}} U_{lk} N_{k}(x),
\end{equation}
with the masses $m_k$ and $M_k$ respectively. Here $U$ is neutrino mixing matrix.

Now let us consider  LNV processes with two charged (anti-)leptons
$({\bar l}_{\alpha})l_{\alpha},
\ ({\bar l}_{\beta})l_{\beta}$ in the initial/final state or with one
$({\bar l}_{\alpha})l_{\alpha}$ in the initial and another
$l_{\beta}, ({\bar l}_{\beta})$ in the final state. Assume that the characteristic
energy scale of this process is $q_0$ and that light and heavy neutrino masses satisfy
the conditions:
\begin{eqnarray}\label{condition}
m_{k} \ll q_0 \ \ \ \mbox{for} \ \ \ \forall k, \ \ \ \mbox{and} \ \ \
M_{k} \gg q_0 \ \ \ \mbox{for} \ \ \ \forall k.
\end{eqnarray}
Then neutrino contribution to its amplitude ${\cal A}_{\alpha\beta}$ can be represented
in the form (for more details see, for instance, Ref. \cite{dib00})
\begin{equation}
  \label{eq.10-1}
  {\cal A}_{\alpha\beta} = \langle m_\nu \rangle_{\alpha\beta}\cdot G_{\nu} +
\langle M_{N}^{-1} \rangle_{\alpha\beta}\cdot G_{N}
\end{equation}
where $G_{\nu}, \ G_N$ are the corresponding structure factors and
\begin{eqnarray}\label{eq.10}
\langle m_\nu \rangle_{\alpha\beta}&=& \sum_{k = \mathrm{light}} U_{\alpha k} U_{\beta k}
 m_k,\\
\label{eq.11}
\ \ \
\langle M_{N}^{-1} \rangle_{\alpha\beta} &=&
\sum_{k=\mathrm{heavy}}\frac{U_{\alpha k} U_{\beta k} }{M_k}
\end{eqnarray}
are the effective light and heavy neutrino masses respectively.

The following comment is in order.  If the mixing of heavy neutrino states
to the active flavors is negligible,
the light neutrino sector can be characterized by the effective light
neutrino mass matrix ${\cal M}^{(\nu)}$ which satisfies the relation
\begin{equation}
  \label{eq.12}
  {\cal M}^{(\nu)}_{\alpha\beta} =
  \langle m_\nu \rangle_{\alpha\beta}.
\end{equation}
If the heavy Majorana neutrino states $N$ are appreciably mixed with the active neutrino
flavors, this equality no longer holds and LNV processes do not provide direct limits
on Majorana neutrino mass matrix elements.

From the non-observation of the LNV processes one can deduce the upper limits on the
corresponding parameters $\langle m_{\nu}\rangle$ and $\langle M_{N}^{  -1}\rangle$.
It must be stressed that these limits have physical sense only if they satisfy the following
consistency conditions
\begin{eqnarray}\label{consistency}
|\langle m_\nu \rangle_{\alpha\beta}|\ll q_0, \ \ \ \
|\langle M_{N}^{-1} \rangle_{\alpha\beta}|^{-1} \gg q_0,
\end{eqnarray}
which follow from the conditions of Eq. (\ref{condition}).

Currently the most stringent limits of this type stem from the $0\nu\beta\beta$-decay.
Its amplitude, written in the form of Eq. (\ref{eq.10-1}), depends on
the parameters $\langle m_{\nu}\rangle_{ee}$ and $\langle M^{-1}\rangle_{ee}$.
Assuming that only light or heavy exchange mechanism is in operation,
the following limits have been derived from the experimental data ~\cite{kla01a,bil04,PaesKlapd}
\begin{eqnarray}\label{nldbd}
|\langle m_\nu \rangle_{ee}| \le 0.55~\mathrm{eV}, \ \ \
\left|\langle M_{N}^{-1}\rangle_{ee}\right|^{-1} \ge 9\times 10^{7}\mbox{GeV}.
\end{eqnarray}
Note that these limits satisfy the consistency conditions in Eq. (\ref{consistency}) since the
characteristic energy scale of $0\nu\beta\beta$-decay is of the order of
$q_0 \sim 100$ MeV.

As we shall demonstrate, the current and near future experimental
searches for $(\mu^-, e^+)$ conversion are unable to reach
meaningful limits on the corresponding parameters $\langle
m_{\nu}\rangle_{\mu e}$ and $\langle M^{-1}\rangle_{\mu e}$
satisfying the consistency conditions in Eq. (\ref{consistency}).
Moreover, the limits following from the neutrino observations and
cosmological data show that the sensitivities of \mue conversion
experiments are too far from being able to detect neutrino
contributions. With the lucky exception of the
$0\nu\beta\beta$-decay this is the fate of all the experiments
searching for other known LNV processes (see, for instance,
\cite{rod02}).

\section{\label{sec:level2-1} Effective neutrino mass from neutrino observations}

Here, we estimate the effective light $\langle m_{\nu}\rangle_{\mu e}$ and heavy
$\langle M_{N}^{-1}\rangle_{\mu e}$ neutrino effective masses
which determine light and heavy Majorana neutrino contributions to $(\mu^{-}-e^{+})$
conversion according to the general formula in Eq. (\ref{eq.10-1}). To this end we  utilize
the existing neutrino oscillation, cosmological and accelerator data, applying the methods
previously used for the analysis of $\langle m_{\nu}\rangle_{ee}$
relevant for $0\nu\beta\beta$-decay
(see, for instance,~\cite{rod03,bil04} and references therein).

Let us start with the three light neutrino scenario without heavy neutrinos.
In this case we have
\begin{equation}
  \label{eq:effmass}
  |\langle m_\nu \rangle_{\mu e}| = \left|
  U_{e1} U_{\mu 1}  m_1 +   U_{e2} U_{\mu 2}   m_2 +  U_{e3} U_{\mu 3}   m_3 \right|,
\end{equation}
with the unitary Pontecorvo-Maki-Nakagawa-Sakata neutrino mixing
matrix $U$. In its standard parametrization (e.g.~\cite{bil99b})
it takes the form
\begin{equation}
  \label{eq:PMNS}
  U =
  \begin{pmatrix}
    c_{12} c_{13}&
    s_{12} c_{13}&
    s_{13}  \mathrm{e}^{- \mathrm{i} \delta}\\
    - s_{12} c_{23} - c_{12} s_{23} s_{13} \mathrm{e}^{\mathrm{i} \delta}&
    c_{12} c_{23} - s_{12} s_{23} s_{13} \mathrm{e}^{\mathrm{i} \delta}&
    s_{23} c_{13}\\
    s_{12} s_{23} - c_{12} c_{23} s_{13} \mathrm{e}^{\mathrm{i} \delta}&
    - c_{12} s_{23} - s_{12} c_{23} s_{13} \mathrm{e}^{\mathrm{i} \delta}&
    c_{23} c_{13}
  \end{pmatrix}
  \begin{pmatrix}
    1& 0& 0\\
    0& \mathrm{e}^{\mathrm{i} \frac{\alpha_{21}}{2}}& 0\\
    0& 0& \mathrm{e}^{\mathrm{i} \frac{\alpha_{31}}{2}}
  \end{pmatrix},
\end{equation}
where $c_{ij} \equiv \cos \theta_{ij}$, $s_{ij} \equiv \sin \theta_{ij}$.
The three mixing angles vary in the range $0 \leq \theta_{ij} \leq \pi/2$.
In addition, Majorana neutrino mixing matrix $U$ contains three CP-violating phases:
one Dirac $\delta$ and two Majorana phases $\alpha_{21}$, $\alpha_{31}$.

The global analysis of the solar, atmospheric, reactor and
accelerator neutrino oscillation data gives the following values
of the neutrino mixing angles~\cite{mal03}:
\begin{align}\label{angles}
  \sin^2 \theta_{12}&= 0.30~ [0.23 - 0.39]\\
  \sin^2 \theta_{13}&= 0.006~ [< 0.054]\\
  \sin^2 \theta_{23}&= 0.52~ [0.31 - 0.72]
\end{align}
and the two independent mass-squared differences
\footnote{Mass-squared difference is defined as $\Delta m^{2}_{ij}
= m^2_i - m^2_j$}.
\begin{align}
  \Delta m^{2}_{sol} = 6.9 \times 10^{-5}~\text{eV}^2
  ~[(5.4 - 9.5) \times 10^{-5}~\text{eV}^2]\\ \label{delta-m}
  \Delta m^{2}_{atm} = 2.6 \times 10^{-3}~\text{eV}^2
  ~[(1.4 - 3.7) \times 10^{-3}~\text{eV}^2]
\end{align}
The values in the square brackets correspond to the $3\sigma$
intervals.

Using the above best values for the neutrino oscillation parameters we
estimate the effective light Majorana neutrino mass
$| \langle m_\nu \rangle_{\mu e}|$ for the three standard cases of
neutrino mass spectrum.

 (1). {\it Normal hierarchy}:  $m_1 \ll m_2 \ll m_3$. In this case
$\Delta m^{2}_{21}\approx \Delta m^{2}_{sol}$, $\Delta m^{2}_{32}\approx \Delta m^{2}_{atm}$
Therefore, one has
\begin{eqnarray}\label{m-norm}
m_1 \ll  \sqrt{\Delta m^{2}_{sol}},\
  m_{2}  \simeq  \sqrt{\Delta m^{2}_{sol}}, \
  m_{3}  \simeq  \sqrt{\Delta m^{2}_{atm}}.
\end{eqnarray}

(2). {\it Inverted hierarchy}: $m_3 \ll m_1 < m_2$. Now,
$\Delta m^{2}_{21}\approx \Delta m^{2}_{sol}$, $\Delta m^{2}_{31}\approx -\Delta m^{2}_{atm}$.
This results in the following estimate for neutrino masses
\begin{eqnarray}\label{m-inv}
m_3 \ll  \sqrt{\Delta m^{2}_{atm}},\
  m_{2}  \simeq  \sqrt{\Delta m^{2}_{atm}}, \
  m_{1}  \simeq  \sqrt{\Delta m^{2}_{atm}}.
\end{eqnarray}

\begin{figure}
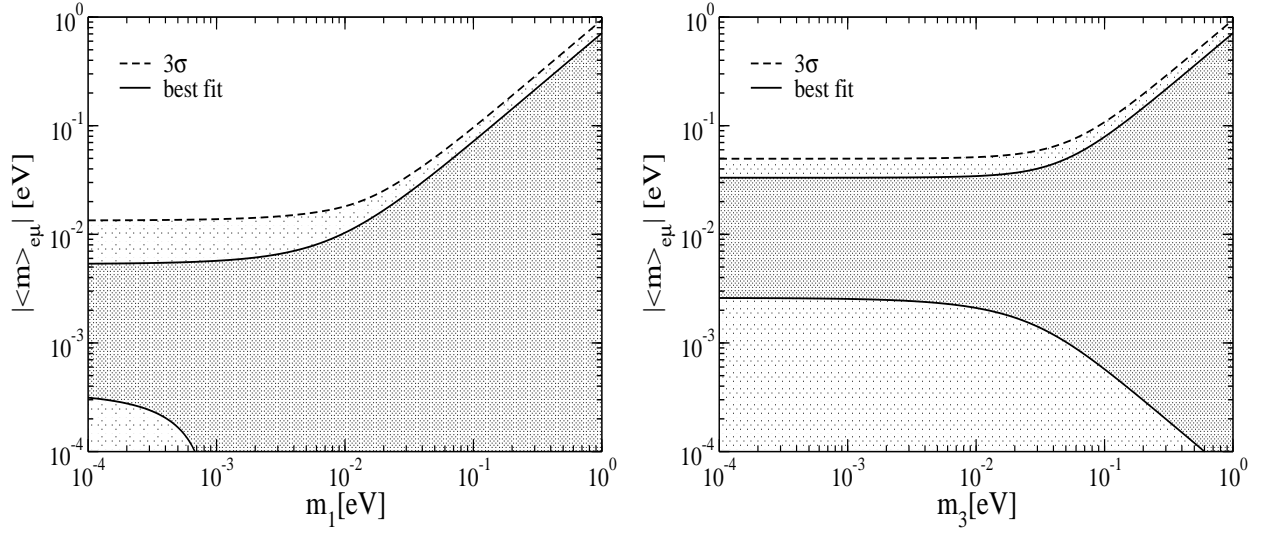

  \begin{center}
    \includegraphics[width=0.49\textwidth,height=0.3\textheight]{normal_black.eps}
    \hfill
    \includegraphics[width=0.49\textwidth,height=0.3\textheight]{inverted_black.eps}
  \end{center}
  \caption{Allowed regions of the effective Majorana neutrino mass $|\langle m
  \rangle_{\mu e}|$ for normal (left panel) and inverted (right panel)
hierarchy vs. the mass of lightest neutrino state: $m_1$ and $m_3$, respectively. \label{fig:regions}}
\end{figure}

Using the estimates (\ref{m-norm})-(\ref{m-inv}) in Eq.~\eqref{eq:effmass} with the best-fit
values for the neutrino oscillation parameters from Eqs.~(\ref{angles})-(\ref{delta-m}),
we end up with the values of the effective light neutrino mass for
\begin{eqnarray}\label{NH}
 &&\mbox{\it Normal hierarchy}:\ |\langle m_\nu \rangle_{\mu e}|
\simeq (0.35 - 5.3) \times 10^{-3}~\text{eV}\\ \label{IH}
&&\mbox{\it Inverted hierarchy}:\ |\langle m_\nu \rangle_{\mu e}|
\simeq (0.3 - 3.3) \times 10^{-2}~\text{eV}.
\end{eqnarray}
within the ranges corresponding to the variation of CP-violating
phases within the intervals $0\leq \delta < 2\pi, \ 0\leq
\alpha_{12} < 2\pi, \ 0\leq \alpha_{23} < 2\pi$. The small terms
with $m_1$ in Eq. (\ref{NH}) and $m_3$ in Eq. (\ref{IH}) were
neglected. The effect of these terms is presented in
Fig.~\ref{fig:regions} which shows the dependence of the allowed
regions of $|\langle m_\nu \rangle_{\mu e}|$ on the mass of the
lightest neutrino $m_1$ for the normal and $m_3$ for the inverted
neutrino mass hierarchies.

(3). {\it Quasi-degenerate hierarchy}: $m_1 \simeq m_2 \simeq m_3$.
This mass spectrum can be consistent with neutrino oscillation data
if the characteristic neutrino mass scale is sufficiently large
$m_0\gg \sqrt{\Delta m^{2}_{atm}}$. In this case the effective light
neutrino mass can be written as
\begin{equation}
  |\langle m_\nu \rangle_{\mu e}| \approx m_0 |\sum_{k=1}^{3} U_{\mu k} U_{e k}|.
\end{equation}
%
In order to estimate its value one needs the values of the characteristic neutrino mass scale
$m_0$. 
It can be deduced from ${^{3}\mathrm{H}}$~experiments and cosmological
data. 
Using the best fit values of neutrino mixing angles from Eq. (\ref{angles}) and
adopting for the simplicity $\delta=\alpha_{12}=\alpha_{23}=0$ we obtain
\begin{align}\label{dg1}
  |\langle m_\nu \rangle_{\mu e}| \lesssim& 1.46~\mathrm{eV}, \quad
  m_0 < 2.05~\mathrm{eV},~\text{Troitsk ${^{3}\mathrm{H}}$
  experiment~\cite{Troitsk}}\\
  |\langle m_\nu \rangle_{\mu e}| \lesssim& 1.56~\mathrm{eV}, \quad
  m_0 < 2.2~\mathrm{eV}, \ \ \, \text{Mainz ${^{3}\mathrm{H}}$
  experiment~\cite{wei03}}\\
  |\langle m_\nu \rangle_{\mu e}| \lesssim& 0.16~\mathrm{eV}, \quad
  m_0 < 0.23~\mathrm{eV},~\text{Cosmological data~\cite{spe03}}\\ \label{dg2}
  |\langle m_\nu \rangle_{\mu e}| \sim& 0.14~\mathrm{eV}, \quad
  m_0 \sim 0.2~\mathrm{eV}, \ \ \, \text{Cosmological data~\cite{all03}}
\end{align}
Note that the results of the global analysis of
the cosmological data in Refs.~\cite{spe03}, \cite{all03} provide significantly more 
stringent limits on the neutrino mass scale than those from the direct laboratory 
measurements of ${^{3}\mathrm{H}}$ $\beta$-decay \cite{Troitsk}, \cite{wei03}.
However, at the same time the cosmological limits are more model dependent than the laboratory ones. 

Now, let us assume that there exist heavy neutrinos $N$ with the masses
$M_k\gg q_0\sim m_{\mu}$, where $q_0\sim m_{\mu}$ is the typical energy
scale of $(\mu^--e^+)$ conversion set by the muon mass $m_{\mu}$.
Their contribution to this process is determined by the effective mass
\begin{eqnarray}\label{eff-N}
\langle M_{N}^{-1} \rangle_{\mu e} =
\sum_{k=\mathrm{heavy}}\frac{U_{\mu k} U_{e k} }{M_k}
\end{eqnarray}
Due to the lack of model independent information on mixing matrix
elements $U_{\mu k} U_{e k}$ in the sector of heavy neutrinos it
is hard to estimate this quantity. For this reason we adopt the
conservative upper bound following from the existing LEP limit on
the mass of heavy stable neutral lepton $M_N\geq 39.5$ GeV
\cite{LEP}. Assuming the existence of only one heavy neutrino
identified with this particle we obtain
\begin{eqnarray}\label{eff-N-LEP}
|\langle M_N^{-1}\rangle_{\mu e}|\leq \left(39.5 \mbox{ GeV}\right)^{-1}.
\end{eqnarray}

In what follows we will use the results presented in Eqs.
(\ref{NH}), (\ref{IH}), (\ref{dg1})-(\ref{dg2}) and
(\ref{eff-N-LEP}) for discussion of the expected rates of
$(\mu^--e^+)$ conversion induced by the Majorana neutrino
exchange.

\section{\label{sec:level3}Neutrino mediated $(\mu^-, e^+)$ conversion. General formalism}

The process of $(\mu^-, e^+)$ conversion is very similar to
the $0\nu\beta\beta$-decay. Both processes violate lepton number
by two units and, therefore, take place if and only if neutrinos are Majorana
particles with non-zero mass.

On the other hand, there are various important differences
between $(\mu^-, e^+)$ conversion and $0\nu\beta\beta$-decay. Among them we mention the following.
\begin{enumerate}[i)]
  \item
    They have rather different available energies and different number of leptons in
    their final states. This results in a significant difference between
    the corresponding phase space integrals.
  \item
    The emitted positron in $(\mu^-, e^+)$ conversion has large momentum and,
    therefore, the long-wave approximation is not valid in
    contrast to $0\nu\beta\beta$-decay.
  \item
    As we will show, the nuclear matrix element of $(\mu^-, e^+)$
    conversion for light neutrino-exchange demonstrates a singular behavior, absent in
    the $0\nu\beta\beta$-decay. This feature gives rise to the large
    imaginary part of the $(\mu^-, e^+)$ conversion amplitude.
    Technically the singularity significantly complicates the numerical calculation of
    the nuclear matrix elements.
  \item
    In the case of the $(\mu^-, e^+)$ conversion there is large number of
    nuclear final states which must be properly taken into account.

\end{enumerate}

\begin{figure}[t]
  \begin{center}
    \includegraphics[height = 12.0cm]{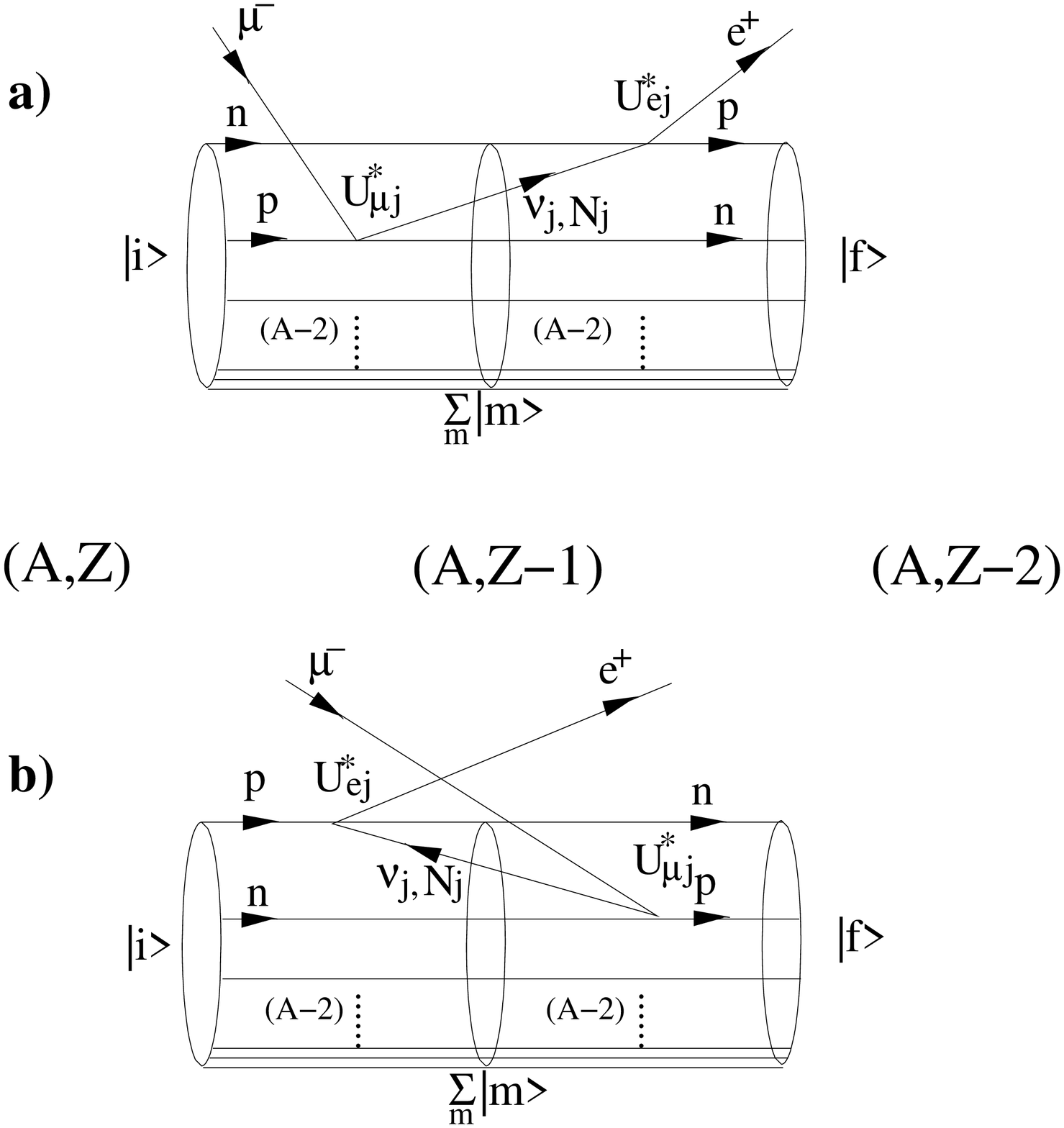}
  \end{center}
  \caption{Direct (a) and cross (b) Feynman diagrams of
  $(\mu^-,e^+)$ conversion in nuclei mediated by Majorana neutrinos.
  \label{fig.1}}
\end{figure}

Below, we analyze the amplitude of the $(\mu^-,e^+)$ conversion in
nuclei mediated by light and heavy Majorana neutrinos. The
corresponding diagrams are shown in Fig.~\ref{fig.1}. We
concentrate only on the nuclear transition connecting the ground
states ($g.s$) of the initial and final nuclei, which is favored
from the experimental point of view due to the minimal background.
The characteristic signature of $g.s\rightarrow g.s.$ transition
is the presence of a peak in the $e^+$ spectrum at the energy
\begin{equation}
  \label{eq.12-1}
  E_{e^+} = m_\mu - \varepsilon_\mathrm{b} - (E_f - E_i)
\end{equation}
which allows reliable separation of signal from background. Here,
$m_\mu$, $\varepsilon_b$, $E_i$ and $E_f$ are the mass of muon,
the muon atomic binding energy (for ${^{48}\mathrm{Ti}}$ this is
$\varepsilon_b = 1.45~\mathrm{MeV}$), the energies of initial and
final nuclear ground states, respectively.  Latter on we neglect
the kinetic energy of final nucleus.
%
%


The leading order $(\mu^-, e^+)$ conversion matrix element, corresponding to the diagrams
in Fig.~\ref{fig.1}, reads
\begin{equation}
  \begin{split}\label{S-matrix}
    \langle f \vert S^{(2)} \vert i \rangle &=
    - \mathrm{i} \left(\frac{G_{\mathrm{F}}}{\sqrt{2}}\right)^2
    \frac{1}{(2\pi)^{3/2}}\frac{1}{\sqrt{4 E_{\mu^-} E_{e^+}}}
    \overline{v}(k_{e^+}) (1 + \gamma_5) u(k_{\mu^-})\\
    &\times \frac{ m_e g^2_{\mathrm{A}}}{2\pi  R }
    \left[\eta^{\mu e}_{\nu} {\cal M}^{(\mu e^{+})\Phi}_{\nu} +
    \eta^{\mu e}_{N} {\cal M}^{(\mu e^{+})\Phi}_{N}\right]
    2 \pi \delta(E_{\mu^-} + E_i - E_f - E_{e^+}).
  \end{split}
\end{equation}
Here $m_e$ and $m_p$ are electron and proton masses, $k_{e^+}$
($E_{e^+}$), $k_{\mu^-}$ ($E_{\mu^-}$) are the momentum (energy)
of outgoing positron and captured muon respectively. The
conventional normalization factor involves the nuclear radius $R =
1.1~A^{1/3}~{\mathrm{fm}}$. For the weak axial coupling constant
$g_A$  we adopt the value $g_A=1.254$. In the above expression we
introduced for convenience the following LNV parameters
\begin{eqnarray}\label{LNV-param}
\eta^{\mu e}_{\nu} = \frac{\langle m_\nu \rangle_{\mu e}}{m_e}, \ \ \
\eta^{\mu e}_{N} = \langle M_{N}^{-1} \rangle_{\mu e} m_p.
\end{eqnarray}

The nuclear matrix elements in Eq. (\ref{S-matrix}) defined as
\begin{equation}
  {\cal M}^{(\mu e^{+})\Phi}_{i} =
  -\frac{M^{(\mu e^{+})\Phi}_{\mathrm{F}({i})}}{g_{\mathrm{A}}^2}
  + M^{(\mu e^{+})\Phi}_{\mathrm{GT}(i)} \ \ \ \ \  \mbox{for} \  \qquad i = \nu, N
\end{equation}
contain the Fermi $M^{(\mu e^{+})\Phi}_{\mathrm{F}}$ and
Gamow-Teller $M^{(\mu e^{+})\Phi}_{\mathrm{GT}}$ contributions.
They take the following form for {\it the light Majorana neutrino
exchange mechanism}
\begin{equation}\label{F-light}
  \begin{split}
    M^{(\mu e^{+})\Phi}_{\mathrm{F}(\nu)} &= \frac{4 \pi  R }{(2 \pi)^3}
    \int \frac{d \vec{q}}{2 q} f^2_{\mathrm{V}}(\vec{q}^{\;2})\\
    &\times \sum_n \left(
    \frac{%
    \langle 0^+_i \vert \sum_l \tau^+_l
    \mathrm{e}^{-\mathrm{i} \vec{k}_{e^+} \cdot \vec{r}_l}
    \mathrm{e}^{-\mathrm{i} \vec{q} \cdot \vec{r}_l}
    \vert n \rangle
    \langle n \vert \sum_m \tau^+_m
    \mathrm{e}^{\mathrm{i} \vec{q}\cdot \vec{r}_m} \Phi (r_m)
    \vert 0^+_f \rangle}
    {q - E_{\mu^-} + E_n - E_i + \mathrm{i} \varepsilon_n} \right.\\
    &+ \left.
    \frac{%
    \langle 0^+_i \vert \sum_m \tau^+_m
    \mathrm{e}^{\mathrm{i} \vec{q} \cdot \vec{r}_m} \Phi (r_m)
    \vert n \rangle \langle n \vert \sum_l \tau^+_l
    \mathrm{e}^{-\mathrm{i} \vec{k}_{e^+} \cdot \vec{r}_l}
    \mathrm{e}^{-\mathrm{i} \vec{q} \cdot \vec{r}_l}
    \vert 0^+_f \rangle}
    {q + E_{e^+} + E_n - E_i + \mathrm{i} \varepsilon_n}
    \right),
  \end{split}
\end{equation}
\begin{equation}\label{GT-light}
  \begin{split}
    M^{(\mu e^{+})\Phi}_{\mathrm{GT}(\nu)} &= \frac{4 \pi  R }{(2 \pi)^3}
    \int \frac{d\vec{q}}{2 q} f^2_{\mathrm{A}}(\vec{q}^{\;2})\\
    &\times \sum_n \left(
    \frac{%
    \langle 0^+_i \vert \sum_l \tau^+_l \vec{\sigma}_l
    \mathrm{e}^{-\mathrm{i} \vec{k}_{e^+} \cdot \vec{r}_l}
    \mathrm{e}^{-\mathrm{i} \vec{q} \cdot \vec{r}_l}
    \vert n \rangle \cdot
    \langle n \vert \sum_m \tau^+_m \vec{\sigma}_m
    \mathrm{e}^{\mathrm{i} \vec{q} \cdot \vec{r}_m} \Phi (r_m)
    \vert 0^+_f \rangle}
    {q - E_{\mu^-} + E_n - E_i + \mathrm{i} \varepsilon_n} \right.\\
    &+ \left.
    \frac{%
    \langle 0^+_i \vert \sum_m \tau^+_m \vec{\sigma}_m
    \mathrm{e}^{\mathrm{i} \vec{q} \cdot \vec{r}_m} \Phi (r_m)
    \vert n \rangle \cdot
    \langle n \vert \sum_l \tau^+_l \vec{\sigma}_l
    \mathrm{e}^{-\mathrm{i} \vec{k}_{e^+} \cdot \vec{r}_l}
    \mathrm{e}^{-\mathrm{i} \vec{q} \cdot \vec{r}_l}
    \vert 0^+_f \rangle}
    {q + E_{e^+} + E_n - E_i + \mathrm{i} \varepsilon_n}
    \right).
  \end{split}
\end{equation}
and for {\it the heavy Majorana neutrino exchange mechanism}
\begin{equation}\label{FGT-heavy}
    M^{(\mu e^{+})\Phi}_{I(N)} = \frac{4 \pi  R }{(2 \pi)^3} \frac{2}{m_p m_e}
    \int {d\vec{q}}
    \langle 0^+_i \vert
    \sum_{lm} \tau^+_l \tau^+_m h_I(\vec{q}^{\;2})
    \mathrm{e}^{-\mathrm{i} \vec{q} \cdot (\vec{r}_l - \vec{r}_m)}
    \mathrm{e}^{-\mathrm{i} \vec{k}_{e^+} \cdot \vec{r}_l} \Phi (r_m)
    \vert 0^+_f \rangle, \, (I=\mathrm{F, GT})
\end{equation}
with
\begin{equation}
  \label{eq:nupotentialFGT}
  h_{\mathrm{F}}(\vec{q}^{\;2}) = f_{\mathrm{V}}^2(\vec{q}^{\;2}),\qquad
  h_{\mathrm{GT}}(\vec{q}^{\;2}) = \vec{\sigma}_{l} \cdot \vec{\sigma}_{m}
  f_{\mathrm{A}}^2(\vec{q}^{\;2}).
\end{equation}
We use the conventional dipole parametrization for the nucleon
form factors~\cite{tow95}
\begin{equation}\label{dipole}
    f_{\mathrm{V}}(\vec{q}^{\;2}) =
    \left(1+\frac{\vec{q}^{\;2}}{\Lambda_{\mathrm{V}}^2} \right)^{-2},
    \qquad
    f_{\mathrm{A}}(\vec{q}^{\;2}) =
    \left(1+\frac{\vec{q}^{\;2}}{\Lambda_{\mathrm{A}}^2} \right)^{-2},
\end{equation}
with $\Lambda_{\mathrm{V}} = 0.71~\mathrm{GeV}$,
$\Lambda_{\mathrm{A}} = 1.09~\mathrm{GeV}$. In Eqs.
(\ref{F-light})-(\ref{FGT-heavy}) the factor $\Phi(r)$ is the
radial part of the bound muon 1S wave function~(see
Appendix~\ref{sec:appA}).
In the denominators of Eqs. (\ref{F-light}), (\ref{GT-light}) we introduced 
the widths $\varepsilon_n$ of intermediate nuclear states.

In the calculations of nuclear matrix elements we adopt the following approximations.
\begin{enumerate}[i)]
  \item
    Taking into account slow variation of muon wave function within the nucleus we
    apply the standard approximation~\cite{kos94}
    \begin{equation}
      |{\cal M}^{(\mu e^{+})\Phi}_{i}|^2 = \langle \Phi \rangle^2
      |{\cal M}_{i}^{(\mu e^{+})}|^2, \quad
      i = \nu, N.
    \end{equation}
    Here $\langle \Phi \rangle^2$ is the muon average probability density and
\begin{eqnarray}\label{approx1}
      \left|{\cal M}_{i}^{(\mu e^{+})}\right| = \left|{\cal M}^{(\mu
      e^{+})\Phi}_{i}\right|_{\Phi=1}\ .
\end{eqnarray}
    The explicit form of $\langle \Phi \rangle^2$ is given in Appendix~\ref{sec:appB}.
  \item
In muon to positron conversion the typical energy of light intermediate neutrinos is
about 100 MeV ($\omega \approx |q| \ge 1/R \sim 100~MeV$) which 
is much larger than the typical excitation energies of intermediate nuclear 
states. 
Therefore, to a good approximation the individual energies of these states
in the energy denominators of Eqs. (\ref{F-light}), (\ref{GT-light}) can 
be neglected or replaced by some average value $<E_n>$ to which  
the matrix elements are not very sensitive.
Then the intermediate nuclear states can be summed up by closure. A similar situation
occurs in the case of $0\nu\beta\beta$-decay \cite{doi85,fae98,suh98}.

    Thus, in Eqs. (\ref{F-light}), (\ref{GT-light}) we complete the sum over 
    the virtual intermediate nuclear states by closure
    after replacing $E_n$, $\varepsilon_n$ with some average values
    $\langle E_n \rangle$, $\varepsilon$, respectively:
    \begin{align}
      \sum_n \frac{\vert n \rangle \langle n \vert}
      {q - E_{\mu^-} + E_n - E_i + \mathrm{i} \varepsilon_n}& 
\approx
      \frac{1}{q - E_{\mu^-} + \langle E_n \rangle - E_i
      + \mathrm{i} \varepsilon},\\
      \sum_n \frac{\vert n \rangle \langle n \vert}
      {q + E_{e^+} + E_n - E_i + \mathrm{i} \varepsilon_n}& 
\approx
      \frac{1}{q + E_{e^+} + \langle E_n \rangle - E_i
      + \mathrm{i} \varepsilon}.
    \end{align}
\end{enumerate}
Obviously, the validity of the closure approximation 
is just the question of the choice of the average excitation energy 
which will be discussed in Section  \ref{sec:level4}.

The angular part of neutrino propagators can be integrated using the relation
\begin{multline}
    \label{anglu}
    \int
    \mathrm{e}^{-\mathrm{i} \vec{q} \cdot (\vec{r}_l - \vec{r}_m)}
    \mathrm{e}^{-\mathrm{i} \vec{k}_{e^+} \cdot \vec{r}_l} d\Omega_q =\\
    (4\pi)^2
    \sum_\lambda (-1)^\lambda \sqrt{2\lambda+1}\,
    \mathrm{j}_\lambda (k_{e^+} R_{lm})
    \mathrm{j}_0(q r_{lm}) \mathrm{j}_\lambda (k_{e^+} r_{lm}/2 )
    \left\{ \mathrm{Y}_\lambda (\Omega_{r_{lm}}) \otimes
    \mathrm{Y}_\lambda (\Omega_{R_{lm}})\right\}_{00},
\end{multline}
Where $\mathrm{j}_\lambda$ is the spherical Bessel function,
$\mathrm{Y}_\lambda$ is the spherical harmonic and
\begin{equation}
    {\vec r}_{ij} = {\vec r}_i-{\vec r}_j, \qquad
    r_{ij}=|{\vec r}_{ij}|, \qquad
    {\vec R}_{ij}=\frac{{\vec r}_i+{\vec r}_j}{2}, \qquad
    R_{ij}=|{\vec R}_{ij}|.
\end{equation}
Note that in the limit when the outgoing positron momentum
$|\vec{k}_{e^+}|$ is zero the right hand side of Eq. (\ref{anglu})
is reduced to $4\pi \mathrm{j}_0(q r_{lm})$.

With the above approximations and comments we can write down the
expressions for the nuclear matrix elements introduced in
Eq.~(\ref{approx1}) in the form
\begin{eqnarray}\label{eq:MEL}
{\cal M}_{\nu}^{(\mu e^{+})} = M_{\mathrm{dir.}}^{(\mu e^{+})} + M_{\mathrm{cro.}}^{(\mu e^{+})},
\ \ \ \ \
{\cal M}_{N}^{(\mu e^{+})} =   -\frac{M_{\mathrm{F}(N)}^{(\mu e^{+})}}{g_{\mathrm{A}}^2} +
  M_{\mathrm{GT}(N)}^{(\mu e^{+})}.
\end{eqnarray}
Here the nuclear matrix element ${\cal M}_{\nu}^{(\mu e^{+})}$ is
decomposed into the contributions coming from direct and cross
Feynman diagrams in Fig~\ref{fig.1}. They can be written as
\begin{equation}
  \label{eq:Mdir}
  \begin{split}
    M_{\mathrm{dir.}}^{(\mu e^{+})} &=
    \langle 0^+_i \vert \sum_{lm} \tau^+_l \tau^+_m
    4\pi \sum_\lambda (-1)^\lambda \sqrt{2\lambda+1}
    \mathrm{j}_\lambda (k_{e^+} R_{lm})
    \mathrm{j}_\lambda \left(\frac{k_{e^+} r_{lm}}{2}\right)
    \left\{ \mathrm{Y}_\lambda (\Omega_{r_{lm}}) \otimes \mathrm{Y}_\lambda
    (\Omega_{R_{lm}})\right\}_{00} \\
    &\times \frac{R}{\pi} \int_0^\infty
    \frac{\mathrm{j}_0(q r_{lm})}
    {q - E_{\mu^-} + \langle E_n \rangle - E_i + \mathrm{i} \varepsilon}
    \left(\vec{\sigma_l} \cdot \vec{\sigma_m} f_{\mathrm{A}}^2(q^2) -
    \frac{f_{\mathrm{V}}^2(q^2)}{g^2_{\mathrm{A}}}\right)
    q dq \vert 0^+_f \rangle,\\
  \end{split}
\end{equation}
\begin{equation}
  \label{eq:Mcro}
  \begin{split}
    M_{\mathrm{cro.}}^{(\mu e^{+})} &=
    \langle 0^+_i \vert \sum_{lm} \tau^+_l \tau^+_m
    4\pi \sum_\lambda (-1)^\lambda \sqrt{2\lambda+1}\,
    \mathrm{j}_\lambda (k_{e^+} R_{lm})
   \mathrm{j}_\lambda \left(\frac{k_{e^+} r_{lm}}{2}\right)
    \left\{ \mathrm{Y}_\lambda (\Omega_{r_{lm}}) \otimes \mathrm{Y}_\lambda
    (\Omega_{R_{lm}}) \right\}_{00} \\
    &\times \frac{R}{\pi} \int_0^\infty
    \frac{\mathrm{j}_0(q r_{lm})}
    {q + E_{e^+} + \langle E_n \rangle - E_i + \mathrm{i} \varepsilon}
    \left(\vec{\sigma_l} \cdot \vec{\sigma_m} f_{\mathrm{A}}^2(q^2) -
    \frac{f_{\mathrm{V}}^2(q^2)}{g^2_{\mathrm{A}}}\right)
    q dq \vert 0^+_f \rangle.
  \end{split}
\end{equation}

The Gamow-Teller and Fermi nuclear matrix elements of heavy Majorana neutrino
exchange mechanism take the form
\begin{equation}
  \label{eq:MN}
  \begin{split}
    M_{I(N)}^{(\mu e^{+})}& = \frac{1}{m_p m_e}
    \langle 0^+_i \vert \sum_{lm} \tau^+_l \tau^+_m
    4\pi \sum_\lambda (-1)^\lambda \sqrt{2\lambda+1}\,
    \mathrm{j}_\lambda (k_{e^+} R_{lm})
    \mathrm{j}_\lambda\left(\frac{k_{e^+} r_{lm}}{2}\right)\\
    &\times
    \left\{ \mathrm{Y}_\lambda (\Omega_{r_{lm}}) \otimes \mathrm{Y}_\lambda
    (\Omega_{R_{lm}}) \right\}_{00}
    \frac{2R}{\pi} \int_0^\infty
    \mathrm{j}_0(q r_{lm}) h_I(\vec{q}^{\;2})
    q^2 dq \vert 0^+_f \rangle \quad (I=\mathrm{F, GT}),
  \end{split}
\end{equation}
with $h_I(\vec{q}^{\;2})$ defined in Eq.~\eqref{eq:nupotentialFGT}.

It is important to note that the value of  
$E_{r} \equiv -E_{\mu^{-}} + \langle E_{n}\rangle - E_{i}$ 
is negative for the studied nuclear
system $A = 48$. Therefore, the contribution of direct Feynman
diagram in Fig.~2a with the light intermediate neutrino has the
pole at 
$q=-E_r-i\varepsilon$, 
as it follows from the formula (\ref{eq:Mdir}).
As a consequence, the imaginary part of the $(\mu^{-}, e^{+})$
conversion amplitude for the case of the light neutrino exchange
can be significant. This fact was first noticed in
Ref.~\cite{sim01a} and then in Refs.~\cite{div01,div02}. In
Ref.~\cite{div02} it was shown that the imaginary part of the
amplitude dominates in the total branching ratio of the $(\mu^{-},
e^{+})$ conversion in ${}^{27}$Al. In the next section we will
demonstrate that the similar conclusion is valid for $(\mu^{-},
e^{+})$ conversion in ${}^{48}$Ti.

The following comment is in order. In the expressions
(\ref{F-light})-(\ref{FGT-heavy}) for nuclear matrix elements
${\cal M}^{(\mu e^{+})\Phi}_{i}$ we neglected the contributions of
the higher order terms of nucleon current (weak-magnetism, induced
pseudoscalar coupling). As suggested by the analogy with
$0\nu\beta\beta$-decay ~\cite{sim99}, these terms should not be
essential for the light neutrino exchange mechanism meanwhile
their contribution in the case of heavy Majorana neutrino exchange
might be significant. However, the detailed study of this effect
is beyond the scope of this paper and will be considered
elsewhere.

Now we are ready to write down the expression for $g.s.\rightarrow
g.s.$ $(\mu^{-}, e^{+})$ conversion rate. For simplicity we assume
that only one mechanism is in operation and present the
corresponding rates for light and heavy Majorana neutrino exchange
mechanisms separately:
\begin{equation}\label{th-rate}
  \Gamma_{i}^{(\mu e^{+})} =
  \frac{1}{\pi} E_{e^+}~ k_{e^+} F(Z-2,E_{e^+}) c_{\mu e}
  \langle \Phi \rangle^2
  |{\cal M}_{i}^{(\mu e^{+})} |^2 |\eta_{i}^{(\mu e)}|^2, \qquad (i = \nu, N)
\end{equation}
where $c_{\mu e} = 2 G_{\mathrm{F}}^4[({m_e m_\mu})/{(4 \pi m_\mu
R )}]^2 g_{\mathrm{A}}^4$, $k_{e^+} = |{\vec k}_{e^+}|$. The
relativistic Coulomb factor $F(Z,E)$ in Eq. (\ref{th-rate}) we
take in the standard form ~\cite{doi85}
\begin{equation}
  \label{eq:Fdoi}
  F(Z,E) = \left[\frac{2}{\Gamma(2\gamma_1 + 1)}\right]^2
  (2 p R)^{2(\gamma_1-1)} |\Gamma(\gamma_1 - \mathrm{i} y)|^2 e^{-\pi y},
\end{equation}
where $\gamma_1 = \sqrt{1-(\alpha Z)^2}$, $\alpha$ is the fine
structure constant, $y = \alpha Z E/p$.

To conclude this section we point out that in our analysis of
$(\mu^-, e^+)$ conversion we limit ourselves by the
$0^+_{g.s.}\rightarrow 0^+_{g.s.}$ transition  which represents a
particular contribution to the total rate of this process. This is
the most favored channel for experimental study since its signal
can be reliably separated from the background as we commented
above.
On the other hand in Ref.~\cite{div02} it was demonstrated that
$0^+_{g.s.}\rightarrow 0^+_{g.s.}$ transition constitutes about
$41\%$ of the total $(\mu^-, e^+)$ conversion rate in $^{27}$Al
and, therefore neglecting the excited final states is a reasonable
approximation. We expect that this conclusion holds for $^{48}$Ti
as well.

\section{\label{sec:level4}Nuclear matrix elements}

We calculate the \mue conversion nuclear matrix elements within the proton-neutron
renormalized Quasiparticle Random Phase Approximation
(pn-RQRPA)~\cite{toi95,sch96,fae97,wod99}. In the present study we focus on
${}^{48}$Ti nucleus utilized as a stopping target in SINDRUM II \cite{doh93}
and PRIME \cite{kunxx} experiments.

\begin{figure}
  \begin{center}
    \includegraphics[width=0.5\textwidth]{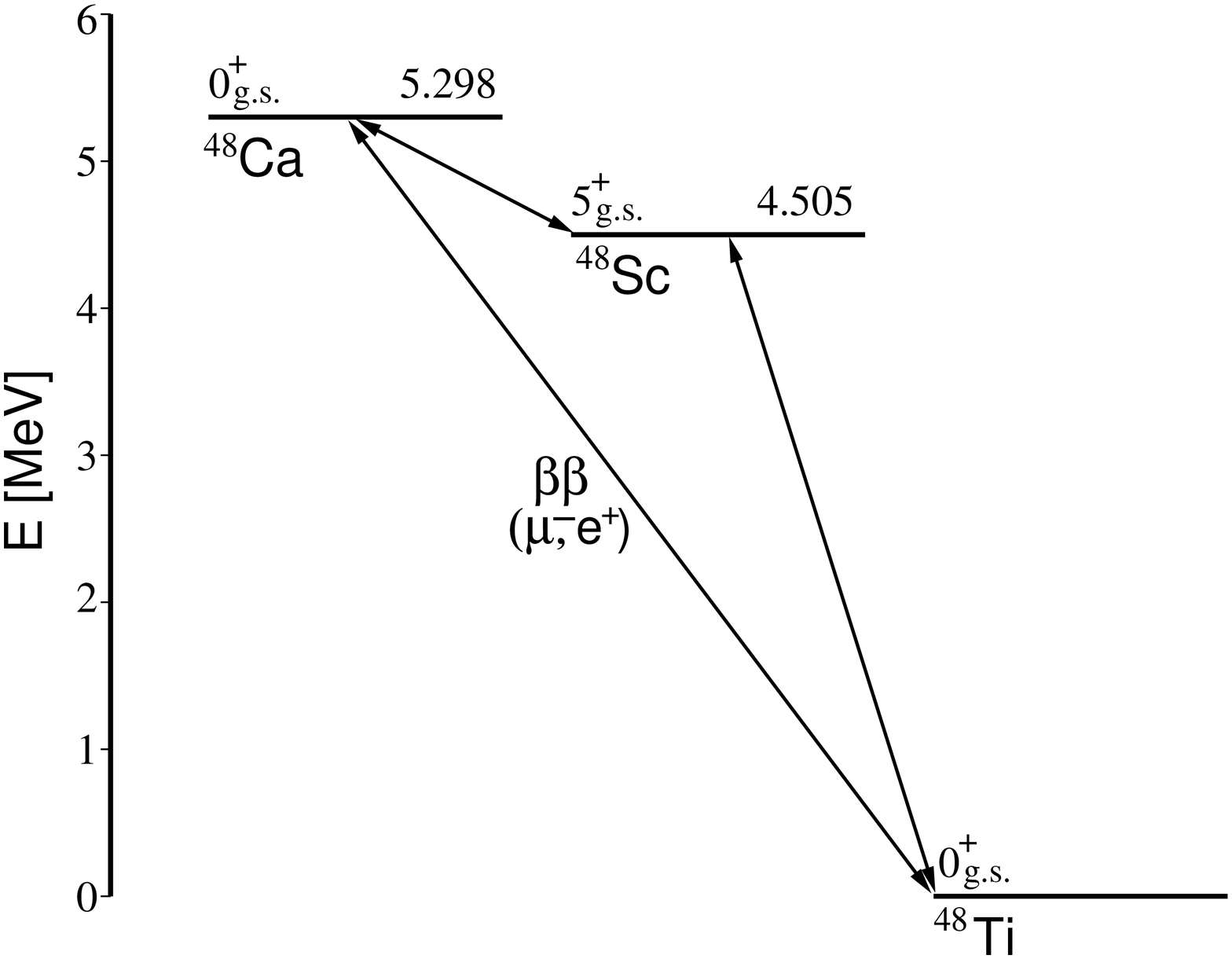}
  \end{center}
  \caption{\label{fig:scheme48ti}Transition scheme for the $A=48$ nuclear system.}
\end{figure}

Nuclear transition scheme for the studied $A=48$ nuclear system is shown
in Fig.~\ref{fig:scheme48ti}
Our nuclear structure calculations involve
the single-particle model space both for protons and neutrons
consisting of the full $0-3\hbar\omega$ shells plus $2s_{1/2}$,
$0g_{7/2}$ and $0g_{9/2}$ levels.  The single particle energies
were obtained using the Coulomb--corrected Woods--Saxon potential.
The two-body G-matrix elements were calculated from the Bonn
one-boson exchange potential on the basis of the Brueckner theory.
Since the considered model space is finite the pairing
interactions have been adjusted to fit the empirical pairing
gaps~\cite{che93a}. In addition, we renormalize the
particle-particle and particle-hole channels of the G-matrix
interaction of the nuclear Hamiltonian $H$ by introducing the
parameters $g_{pp}$ and $g_{ph}$, respectively. The two-nucleon
correlation effect has been taken into account in a standard way
by multiplying the operators with the square of the correlation
Jastrow-like function \cite{jastr}. The details of our nuclear
model  can be found in Appendix~\ref{sec:AppC}

As we already commented in section \ref{sec:level3} the  matrix
element  of the direct contribution (Fig. 2a)
of light neutrino exchange mechanism contains an imaginary part which
stems from the pole of the integrand in Eq. (\ref{eq:Mdir}) 
at $q=-E_r-i\varepsilon$.
Taking into account that the widths $\varepsilon$ of low lying nuclear states are negligible
in comparison with their energies one can separate the imaginary and real parts of this matrix
element using the well known formula
\begin{equation}
  \frac{1}{\alpha + \mathrm{i}\varepsilon} =
  {\cal{P}} \frac{1}{\alpha} - \mathrm{i} \pi  \delta(\alpha)
\end{equation}
valid in the limit $\varepsilon \rightarrow 0$.

\begin{table}[t]
  \caption{Nuclear matrix elements of the light and heavy Majorana neutrino
  exchange mechanisms of $(\mu^-,e^+)$ conversion in $^{48}$Ti [see Eqs.
  \eqref{eq:MEL}-\eqref{eq:MN}].  The
  calculations have been performed within pn-RQRPA without and with
  the inclusion of two-nucleon short-range correlations
  (s.r.c.).\label{table.1}}
  \begin{tabular}{c@{\hspace{10mm}}c@{\hspace{5mm}}cc@{\hspace{5mm}}c@{\hspace{20mm}}r}
    \hline
    \hline
    $g_{pp}$&
    $M_{\mathrm{cro.}}^{(\mu e^{+})}$&
    $\mathrm{Re}(M_{\mathrm{dir.}}^{(\mu e^{+})})$&
    $\mathrm{Im}(M_{\mathrm{dir.}}^{(\mu e^{+})})$&
    $|{\cal M}_{\nu}^{(\mu e^{+})}|$&
    $|{\cal M}_{N}^{(\mu e^{+})}|$\\
    \hline
    \multicolumn{6}{c}{without s.r.c.}\\
    0.8& 0.097&  0.002&  0.088&  0.132& 25.5\\
    1.0& 0.077&  0.034&  0.059&  0.125& 22.8\\
    1.2& 0.051&  0.091&  0.018&  0.142& 19.6\\
    \multicolumn{6}{c}{with s.r.c.}\\
    0.8& 0.049& -0.080&  0.050&  0.059& 5.92\\
    1.0& 0.034& -0.040&  0.024&  0.025& 5.19\\
    1.2& 0.013&  0.027& -0.013&  0.042& 4.33\\
    \multicolumn{6}{c}{with s.r.c., $|\vec{k}_{e^+}| = 0$ }\\
    0.8& 0.298& -0.029&  0.386&  0.470& 31.4\\
    1.0& 0.233&  0.069&  0.275&  0.408& 27.7\\
    1.2& 0.147&  0.243&  0.125&  0.409& 23.2\\
    \hline
    \hline
  \end{tabular}
\end{table}

\begin{figure}[t]
  \begin{center}
    \includegraphics[height = 10.0cm]{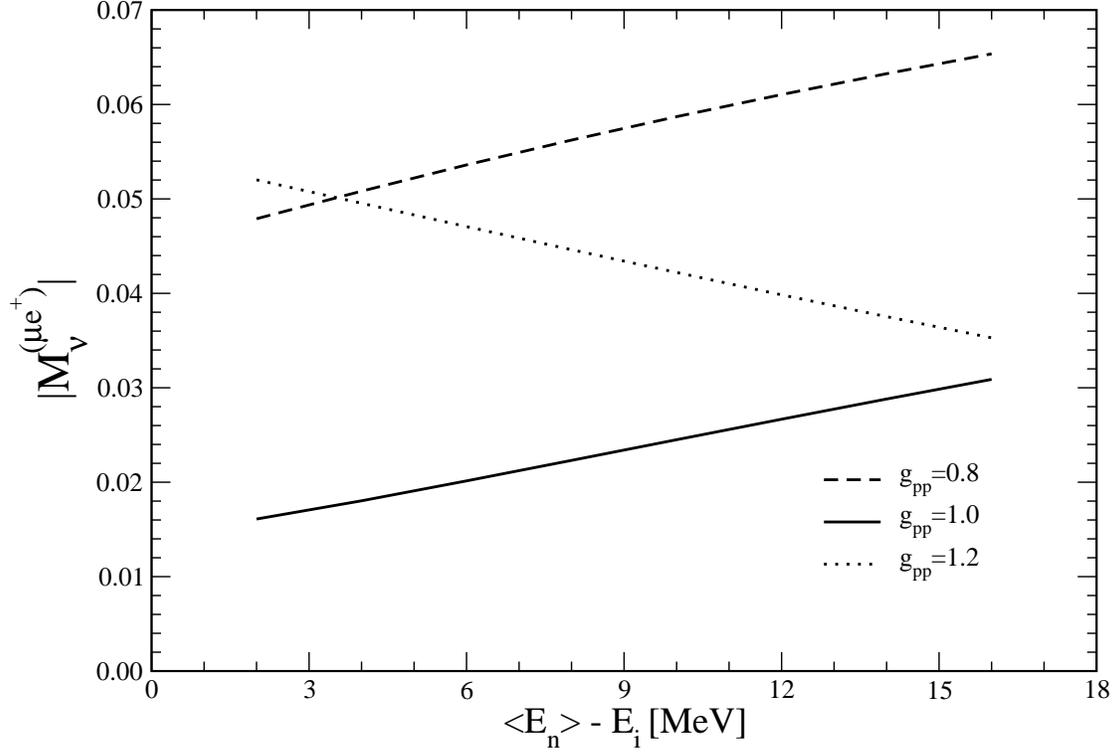}
  \end{center}
  \caption{
  The nuclear matrix elements of the light Majorana neutrino
  exchange mechanisms of the $(\mu^-,e^+)$ conversion in $^{48}Ti$ 
  as a function of the average value of energy difference $<E_n>-E_i$.
  \label{fig.aver}}
\end{figure}

In Table~\ref{table.1} we show the nuclear matrix elements of
light and heavy Majorana neutrino exchange mechanisms of the
$(\mu^-,e^+)$ conversion in ${^{48}\mathrm{Ti}}$ calculated for
$g_{pp} = 1.0$ and $g_{pp} = 0.8,~1.0,~1.2$. All the presented
results were obtained for the particular value of energy
difference 
$\langle E_n \rangle - E_i = 10~\mathrm{MeV}$. 
This choice is justified by weak dependence of the matrix elements on 
this parameter within the interval of its reasonable values 
$2~\mathrm{MeV} \le ( \langle E_n \rangle - E_i) \le 15~\mathrm{MeV}$.
We verified this property by the direct numerical analysis. 
In Fig. \ref{fig.aver} we present the absolute value of the light neutrino exchange nuclear matrix element 
$|M^{(\mu e^{+})}|$ as a function of the average value $\langle E_n \rangle-E_i$ for $g_{pp} = 0.8$, 1.0 and
1.2. One can see that its variation within the studied range
of $\langle E_n \rangle-E_i$ is about 30\%. For $g_{pp}=0.8$, $1.0$ ($g_{pp}=1.2$) the
matrix element is an increasing (decreasing) function of $\langle E_n \rangle-E_i$. 
Different behavior in these two cases is related to a specific interplay between  
the direct $M_{\mathrm{dir.}}^{(\mu e^{+})}$ and cross 
$M_{\mathrm{cro.}}^{(\mu e^{+})}$ diagram terms in
$M^{(\mu e^{+})}$. For $g_{pp}=0.8$, $1.0$ there is a mutual cancellation of 
the real parts of these two terms so that the imaginary part of $M_{\mathrm{dir.}}^{(\mu e^{+})}$,
which is a growing function of  $\langle E_n \rangle-E_i$, dominantes and determines the behavior of $M^{(\mu e^{+})}$.
For $g_{pp}=1.2$ the situation is opposite. The real parts, decreasing with $\langle E_n \rangle-E_i$, contribute coherently
and constitute the dominant part of $M^{(\mu e^{+})}$ which becomes a decreasing function of $\langle E_n \rangle-E_i$.   
%
%
%
We have also found that the nuclear matrix elements do not show an
appreciable variation in the physical region of the parameter
$g_{ph}$ ($0.8 \le g_{ph} \le 1.2$). On the contrary, as seen from
Table~\ref{table.1} they significantly depend on the
renormalization parameter $g_{pp}$ and on the two-nucleon
short-range correlation.
It is also worth noting that the large momentum $k_{e^+}$ of
outgoing positron is the source of strong suppression of the
$(\mu^-,e^+)$ conversion matrix elements. In order to illustrate
this effect we presented in Table~\ref{table.1} the matrix
elements calculated in the limit $|k_{e^+}|=0$ when the
suppression of this type is absent. The cross check of
Table~\ref{table.1} reveals the corresponding suppression factor
of about $\sim 10$.

An important issue of our analysis is the presence of the significant imaginary part of
matrix element ${\cal M}_{\nu}^{\mu e^{+}}$ corresponding to the light Majorana neutrino
exchange mechanism. This fact was first noticed in Ref.~\cite{sim01a}
and then in Refs.~\cite{div01,div02}. In the previous studies of $(\mu^-,e^+)$
conversion \cite{doi85,kam79,ver81,leo83} the role of imaginary part was overlooked.

In the presented detailed study we have found, that the relative
contribution of the imaginary part to the rate of $(\mu^-,e^+)$
conversion in ${}^{48}$Ti is always significant but appreciably
depends on the value of the nuclear model parameter $g_{pp}$ and
on the short range correlations.
It absolutely dominates over the real part by the factor of $\sim
16$ for the most conventional case when $g_{pp}=1$ and the short
range correlations are taken into account (for the motivation of
this choice see, for instance, Ref.~\cite{rod03,bil04}).
This conclusion is consistent with the result of Ref.~\cite{div02}
studying $(\mu^-,e^+)$ conversion in ${^{27}\mathrm{Al}}$ within
 shell-model approach where it was found that the imaginary part
for the light neutrino exchange dominates over the real one by the
factor of about~$~20$. However it is notable that the relative
contribution of the imaginary part is model dependent and can vary
from one nucleus to another. In this situation the role of the
imaginary part in $(\mu^-,e^+)$ conversion requires further study
for other nuclear systems.

%

From the view point of nuclear structure theory it is instructive
to compare the values of \mue conversion nuclear matrix elements
with the corresponding values of $0\nu\beta\beta$-decay matrix
elements of $\mathrm A = 48$ nuclear system. For
$0\nu\beta\beta$-decay this system is represented by $^{48}$Ca
with the matrix elements
\begin{equation}
  |{\cal M}_{\nu}^{(ee)}| = 0.82, \quad
|{\cal M}_{N}^{(ee)}| = 24.2 \label{eq.25}
\end{equation}
derived within the pn-RQRPA approach in Ref.~{\cite{sim01c}.
As seen, the matrix elements of the $(\mu^-,e^+)$ conversion
(\ref{eq:usedMEL}) are strongly suppressed in comparison with
those of $0\nu\beta\beta$-decay (\ref{eq.25}) by the factors of
about $17$ and ~$5$ for the light and heavy Majorana neutrino
exchange mechanisms respectively. As we commented above the
explanation of this difference between the two processes mostly
resides in the large momentum of outgoing positron produced in
$(\mu^-, e^+)$ conversion.

\section{\label{sec:level4-1}$(\mu^-, e^+)$ conversion and effective neutrino masses}

Now, let us discuss the possible issues of  \mue conversion experiments for neutrino physics.

From Eq. (\ref{th-rate}) we obtain the \mue conversion branching
ratios in ${}^{48}$Ti for the light and heavy Majorana neutrino
exchange mechanisms:
\begin{equation}\label{b-rate}
  R^{(\mu e^{+})}_{i} \equiv
  \frac{\Gamma_{i}^{(\mu e)}}{\Gamma^{(\mu \nu)}} = 2.6 \times
  10^{-22}
  |{\cal M}^{(\mu e)}_{i}|^2 |\eta_{i}^{(\mu e)}|^2 \quad (i = \nu, N).
\end{equation}
Here we use the known experimental value $\Gamma^{(\mu \nu)} =
2.60 \times 10^6~\mathrm{s}^{-1}$~\cite{suz87} of ordinary muon
capture rate in ${}^{48}$Ti. For the further discussion we choose
the following sample values of nuclear matrix elements of
$^{48}$Ti from Table~\ref{table.1}
\begin{equation}
  \label{eq:usedMEL}
  |{\cal M}_{\nu}^{(\mu e^{+})}| = 0.025, \quad
  |{\cal M}_{N}^{(\mu e^{+})}| = 5.2
\end{equation}
corresponding to $g_{pp}=1.0$ with the presence of the two-nucleon
short range correlations.

Substituting these numerical values of nuclear matrix elements to
Eq.~(\ref{b-rate}) we obtain
\begin{align}\label{light-r}
  R^{(\mu e^{+})}_{\nu}&
  = 1.6~10^{-25} \times \frac{|{\langle m \rangle_{\mu e}}|^2}{m_e^2},\\
\label{heavy-r}
  R^{(\mu e^{+})}_{N}&
  = 7.0~10^{-21} \times |\langle M_{N}^{-1} \rangle_{\mu e}|^2 m_p^2.
\end{align}
From the existing experimental upper bound in Eq. (\ref{eq.2}) one obtains the following
limits for the effective masses of light and heavy Majorana neutrinos
\begin{eqnarray}\label{effmass-limits}
|{\langle m \rangle_{\mu e}}|\leq 1.3\times 10^{6}~\mbox{MeV} \ \ \ \ \ \ \ \ \ \
|\langle M_{N}^{-1} \rangle_{\mu e}|^{-1} \geq 3.3\times 10^{-2}~\mbox{MeV}
\end{eqnarray}
Obviously, these limits have no physical sense since they do not
satisfy the consistency condition in Eq. (\ref{consistency}) with
the characteristic energy scale $q_0 \sim m_{\mu}=105$MeV of \mue
conversion. Meaningful limits on the parameters ${\langle m
\rangle_{\mu e}}$, $\langle M_{N}^{-1} \rangle_{\mu e}$, which may
have some impact on neutrino physics, could be reached if the \mue
conversion experiments would improve their sensitivities by at
least 10 orders of magnitude. Clearly, such a tremendous
improvement is unrealistic for the near future experiments.


On the other hand we can estimate the expected branching ratios of
\mue conversion induced by the light and heavy Majorana neutrino
exchange using the estimates of ${\langle m \rangle_{\mu e}}$,
$\langle M_{N}^{-1} \rangle_{\mu e}$ made in section
\ref{sec:level2-1} from the present neutrino data. Substituting
the values of these parameters in Eqs.
(\ref{light-r})-(\ref{heavy-r}) we obtain the following results.

\noindent
%
{\it Light Majorana neutrino exchange contribution}:
\begin{enumerate}[i)]
  \item
    Normal neutrino mass hierarchy, $|<m>_{\mu e}| \simeq (0.35 - 5.3) \times
    10^{-3}~\text{eV}$
    \begin{equation}\label{light-lim-norm}
      R^{(\mu e^{+})}_{\nu} \simeq (0.008 - 1.7)\times 10^{-41}.
    \end{equation}
  \item
    Inverted neutrino mass hierarchy, $|<m>_{\mu e}| \simeq (0.3 - 3.3) \times
    10^{-2}~\text{eV}$
    \begin{equation}
      R^{(\mu e^{+})}_{\nu} \simeq (0.05 - 6.7) \times 10^{-40}.
    \end{equation}
  \item
    Quasidegenerate mass hierarchy
  \begin{align}
\label{H-1}
    R^{(\mu e^{+})}_{\nu} \lesssim& 1.3 \times 10^{-36}, \quad
    \langle m_\nu \rangle < 1.46~\mathrm{eV},~\text{Troitsk ${^{3}\mathrm{H}}$
    experiment~\cite{Troitsk}}\\
\label{H-2}
    R^{(\mu e^{+})}_{\nu} \lesssim& 1.5 \times 10^{-36}, \quad
    \langle m_\nu \rangle < 1.56~\mathrm{eV},~\text{Mainz ${^{3}\mathrm{H}}$
    experiment~\cite{wei03}}\\
\label{C-1}
    R^{(\mu e^{+})}_{\nu} \lesssim& 1.6 \times 10^{-38}, \quad
    \langle m_\nu \rangle < 0.16~\mathrm{eV},~\text{Cosmological data~\cite{spe03}}\\
\label{C-2}
    R^{(\mu e^{+})}_{\nu} \sim& 1.3 \times 10^{-38}, \quad
    \langle m_\nu \rangle \sim 0.14~\mathrm{eV},~\text{Cosmological data~\cite{all03}}
  \end{align}
\end{enumerate}
Let us remind that the cosmological data based limits (\ref{C-1})and (\ref{C-2}), albeit more stringent, 
are more model dependent than the laboratory ones (\ref{H-1}) and (\ref{H-2}). 

\noindent
{\it Heavy Majorana neutrino contribution}:
\begin{eqnarray}\label{heavy-limit}
R^{(\mu e^{+})}_{N} \leq  3.8 \times 10^{-24}
\end{eqnarray}

All the values of \mue conversion branching ratio in Eq.
(\ref{light-lim-norm})-(\ref{heavy-limit}) are hopelessly low for
being detected even in a distant future. Thus, searching for \mue
conversion cannot have any direct impact on neutrino physics. On
the other hand any observation of \mue conversion at branching
ratios above the limits in Eq.
(\ref{light-lim-norm})-(\ref{heavy-limit}) would be unambiguous
signal of new physics beyond the simplest extension of SM with
massive Majorana neutrinos and would imply the presence of new
interactions.

This conclusion are in a sharp contrast with $0\nu\beta\beta$-decay experiments which
already provide an important information on neutrino properties and are expected to detect
neutrino contribution in the near future. This is due to their unique sensitivities to $0\nu\beta\beta$-decay
signal. In order to give an impression to which extent $0\nu\beta\beta$-decay experiments
overcome in sensitivities the experiments searching for \mue conversion
let us compare, as an example, the rates of $(\mu^-, e^+)$ conversion in
${}^{48}$Ti and  $0\nu\beta\beta$-decay of ${^{48}\mathrm{Ca}}$.
To this end it is sufficient to consider only light Majorana neutrino
exchange contributions in both cases.
For the rate of $0\nu\beta\beta$-decay we have the well known formula
\begin{equation}
  \Gamma^{(ee)}_{\nu} = \ln 2 G_{01}
  \left|\frac{\langle m_\nu \rangle_{e e}}{m_e}\right|^{2} |{\cal M}^{(ee)}_{\nu}|^2,
\end{equation}
where $G_{01} =
8.031\times 10^{-14}~\mathrm{year}^{-1}$~\cite{pan96} and
\begin{equation}
  \langle m_\nu \rangle_{e e} = \sum\limits_{k=\mathrm{light}}
  (U_{e k})^2  m_k.
\end{equation}
Using the value of $0\nu\beta\beta$-decay nuclear matrix element
${\cal M}^{(ee)}_{\nu}$  from Eq. (\ref{eq.25}) we estimate the ratio
of \mue conversion to $0\nu\beta\beta$-decay rates:
\begin{equation}
  \frac{\Gamma_{\nu}^{(\mu e^{+})}}{\Gamma_{\nu}^{(ee)}} = 9.7 \times10^{4}
  \times \frac{|{\cal M}^{(\mu e)}_{\nu}|^2}{|{\cal M}^{(ee)}_{\nu}|^2}
  \left| \frac{\langle m_\nu \rangle_{\mu e}}{\langle m_\nu \rangle_{ee}}
  \right|^2 =
  351 \left| \frac{\langle m_\nu \rangle_{\mu e}}{\langle m_\nu \rangle_{ee}}
  \right|^2.
\end{equation}
The the $(\mu^-,e^+)$ conversion receives a significant
enhancement mostly due to the larger available energy of this
process. Thus, for ${\langle m_\nu \rangle_{\mu e}} \sim \langle
m_\nu \rangle_{ee}$ the \mue conversion rate ${\Gamma_{\nu}^{(\mu
e^+)}}$ is by more than 2 orders of magnitude larger than the rate
${\Gamma_{\nu}^{(ee)}}$ of $0\nu\beta\beta$-decay. Nevertheless
the experimental prospects for searching for
$0\nu\beta\beta$-decay are incomparably better than those for \mue
conversion. This is mainly because the number of potentially
$0\nu\beta\beta$-decaying nuclei monitored in $0\nu\beta\beta$
experiments is by many orders of magnitude larger than the number
of mesoatoms created by muon beams in the muon-conversion
experiments.


\section{\label{sec:level5}Summary and outlook}

In summary, the light and heavy Majorana neutrino exchange mechanisms
of $(\mu^-,e^+)$ conversion have been studied.
Special emphasis was made on the nuclear structure aspects of this process.
We have performed the first realistic calculations of the
corresponding nuclear matrix elements for ${}^{48}$Ti nucleus used
as a stopping target in the current \cite{doh93} and the
forthcoming \cite{kunxx} \mue conversion experiments. Our analysis
is based on the pn-RQRPA approach and limited to the case of
$0^+_{g.s.} \rightarrow 0^+_{g.s.}$ transition channel, which is
most relevant for experimental searches for \mue conversion. The
effects of the ground state and two-nucleon short-range
correlations have been properly taken into account. We pointed out
that their inclusion results in the significant reduction of
$(\mu^-,e^+)$ conversion matrix elements.

Our detailed analysis confirmed the conjecture of Refs.
\cite{sim01a,div01} on the importance of the imaginary part of the
nuclear matrix elements for the case of the light Majorana
neutrino exchange mechanism of \mue conversion. The similar result
was recently obtained in Ref. \cite{div02} for \mue conversion in
${}^{27}$Al.

We also derived the limits on the effective masses of light
${\langle m \rangle_{\mu e}}$ and heavy $\langle M_{N}^{-1}
\rangle_{\mu e}$ Majorana neutrinos from the neutrino
oscillations, tritium beta decay, accelerator and cosmological
data. Using these limits we estimated the expected rates of \mue
conversion induced by Majorana neutrino exchange. Their values
were found to be so small that even within a quite distant future
the \mue conversion experiments will  hardly be able to detect the
neutrino contribution and, thus, to have a direct impact on
neutrino physics. On the other hand the eventual observation of
\mue conversion at larger rates would be unambiguous signal of new
physics beyond the standard model implying new non-standard
interactions. Moreover, this observation, independently of the
\mue conversion rate, would definitely prove that neutrinos are
Majorana particles as follows from the ``black box" type
theorem~\cite{sch82} establishing the fundamental relation between
LNV processes and Majorana nature of neutrinos. In view of this it
remains actual to study possible scenarios of new physics
consistent with the values of \mue conversion rates within the
reach of the present and near future experiments.

\begin{acknowledgments}
We are grateful to I. Schmidt for useful comments and remarks.
This work was supported in part by Fondecyt (Chile) under
grant 1030244,  by the DFG (Germany) under contract 436 SLK 113/8
and by the VEGA Grant agency
of the Slovak Republic under contract No.~1/0249/03.
\end{acknowledgments}

\appendix
\section{\label{sec:appA}Bound muon wave-function}

The bound muon wave function (1S-state) is given by the expression
\begin{equation}
  \Psi (x) = \Phi (r) e^{- \mathrm{i} E_{\mu^-} x_0}
  \frac{u_\mu^s}{\sqrt{2 E_{\mu^-}}},
\end{equation}
where the radial $\Phi (r)$ and the spinorial $u_\mu^s$ parts have
the forms
\begin{equation}
  \Phi (r) = \frac{Z^{3/2}}{(\pi a_\mu^3)^{1/2}}
  e^{- Z r /a_\mu}
\end{equation}
and
\begin{equation}
  u^s_\mu = \sqrt{2 E_{\mu^-}}
  \begin{pmatrix}
    \chi^s\\
    0
  \end{pmatrix},
\end{equation}
with $a_\mu = 4\pi/(m_\mu e^2)$
($a_\mu/a_e \approx m_e/m_\mu \approx 5\times 10^{-3}$),
$m_\mu$ is reduced mass of muon atom, $Z$ is nuclear charge.

\section{\label{sec:appB}Muon average probability density over nucleus}

Muon average probability density over nucleus is defined as
\begin{equation}
  \langle \Phi \rangle^2 \equiv
  \frac{\int |\Phi(\vec{x})|^2 \rho(\vec{x}) d^3x}{%
  \int \rho(\vec{x}) d^3 x},
\end{equation}
where $\rho(\vec{x})$ is the nuclear charge density. To a good
approximation it can be written in the following compact
form~\cite{kos94}
\begin{equation}
  \langle \Phi \rangle^2 = \frac{\alpha^3 m_{\mu}^3}{\pi}
  \frac{Z_{\mathrm{eff}}^4}{Z}.
\end{equation}
Here the effective charge for $Z = 22$ nuclear system
is $Z_{\mathrm{eff}}$ is $Z_{\mathrm{eff}} = 17.5$ ~\cite{kos94}.

\section{\label{sec:AppC} Nuclear Model}

Here we shortly outline our approach to the nuclear structure
calculations.

We introduce particle (quasiparticle) creation operators as
$c^{\dagger}_{\tau m_{\tau}}$ ($a^{\dagger}_{\tau m_{\tau}}$) for
$\tau = p, n$. The indices $p \equiv (n_p,l_p,j_p)$ and $n \equiv
(n_n,l_n,j_n)$ denote proton and neutron quantum numbers in a
particular shell. Transformation from the particle to
quasiparticle basis is realized by the Bogolyubov transformation
\begin{equation}
  \begin{pmatrix}
    c^{\dagger}_{\tau m_{\tau}}\\
    \tilde{c}_{\tau m_{\tau}}
  \end{pmatrix} =
  \begin{pmatrix}
    u_{\tau}& -v_{\tau}\\
    v_{\tau}& u_{\tau}
  \end{pmatrix}
  \begin{pmatrix}
    a^{\dagger}_{\tau m_{\tau}}\\
    \tilde{a}_{\tau m_{\tau}}
  \end{pmatrix}
\end{equation}
where the tilde denotes time reversal, $\tilde{a}_{\tau m_{\tau}} =
(-1)^{j_{\tau}-m_{\tau}} a_{\tau\,-m_{\tau}}$.

Occupation amplitudes $u_{\tau}$, $v_{\tau}$ and quasiparticle energies
$E_{\tau}$ are obtained by solving BCS equation~\cite{che93b}
\begin{equation}
  \begin{pmatrix}
    \varepsilon_{\tau} - \lambda_{\tau}& \Delta_{\tau}\\
    \Delta_{\tau}& -\varepsilon_{\tau} + \lambda_{\tau}
  \end{pmatrix}
  \begin{pmatrix}
    u_{\tau}\\
    v_{\tau}
  \end{pmatrix} =
  E_{\tau}
  \begin{pmatrix}
    u_{\tau}\\
    v_{\tau}
  \end{pmatrix},
  \label{eq:BCS}
\end{equation}
where $\varepsilon_{\tau}$ is the energy of single particle state
derived from the Wood--Saxon potential.
The pairing potential takes the form
\begin{equation}
  \Delta_{\tau} = (2 j_{\tau} + 1)^{-1/2} \sum_{a} (2 j_a + 1)^{1/2}
  G(a a,\tau \tau ; J=0) u_a v_a.
\end{equation}
Here $G(a a, \tau \tau ; J)$ is particle-particle matrix element
defined {\it e.g.} in Ref.~\cite{r&s80}. The value of Lagrange
multiplier $\lambda$ is fixed by
the particle number $N$ in non-correlated BCS vacuum
\begin{equation}
  \langle N_{\tau} \rangle = \sum\limits_{\tau} (2j_{\tau} + 1) v_{\tau}^2
\end{equation}
After the diagonalization, BCS Eq.\eqref{eq:BCS} reads
\begin{align}
  E_{\tau} = \sqrt{(\varepsilon_{\tau}
  - \lambda_{\tau})^2
  + \Delta^2_{\tau}},~
  v^2_{\tau} = \frac{1}{2}(1-\frac{\varepsilon_\tau - \lambda_\tau}{E_\tau}),
  ~ u^2_\tau = 1 - v^2_\tau.
  \label{eq:BCS-diag}
\end{align}
This system of equations can be solved by the iteration of the parameter
$\lambda_{\tau}$ with the condition $N = \langle N_{\tau} \rangle$.

The nuclear Hamiltonian in quasiparticle representation takes after the BCS
transformation the form
\begin{equation}
  H = \sum\limits_{\tau m_{\tau}}
  E_{\tau} a^+_{\tau m_{\tau}} a_{\tau m_{\tau}} + H_{22} + H_{40} + H_{04} +
  H_{31} + H_{13},
\end{equation}
where $H_{ij}$ is the normally ordered part of residual
interaction with $i$ creation and $j$ annihilation operators.

Within pn-RQRPA, the $m$-th nuclear excited state $\vert m,JM \rangle$
with the angular momentum $J$ and its projection $M$
is obtained from the RPA vacuum $\vert 0^+_{\mathrm{RPA}} \rangle$
\begin{equation}
  \vert m, JM \rangle = Q^{m\dagger}_{JM^{\pi}} \vert 0^+_{\mathrm{RPA}}
  \rangle,
\end{equation}
where RPA vacuum is defined by the condition
\begin{equation}
  Q^{m}_{JM^{\pi}} \vert 0^{+}_{\mathrm{RPA}} \rangle = 0.
\end{equation}
and phonon operator $Q^{m}_{JM^{\pi}}$ is defined as
\begin{equation}
  Q^{m\dagger}_{JM^{\pi}} = \sum\limits_{pn} X^m_{(pn,J^{\pi})}
  A^{\dagger}_{(pn,JM)} - Y^m_{(pn,J^{\pi})} \tilde{A}_{(pn,JM)},
\end{equation}
where $A^{\dagger}_{(pn,J^{\pi})}$ ($\tilde{A}_{(pn,J^{\pi})}$) is
two-particle creation (annihilation) operator which couples
quasiparticles to the angular momentum $J$ with the projection
$M$:
\begin{equation}
  A^{\dagger}(pn,JM) = \sum\limits_{m_p,m_n}
  \mathrm{C}_{j_p m_p j_n m_n}^{JM}
  a^{\dagger}_{p m_p} a^{\dagger}_{n m_n},
\end{equation}
\begin{equation}
  \tilde{A}(pn,JM) = (-1)^{J-M} A(pn,JM) = (-1)^{J-M}
  \sum\limits_{m_p,m_n} \mathrm{C}_{j_p m_p j_n m_n}^{J -M}
  a_{p m_p} a_{n m_n}.
\end{equation}
Here $\mathrm{C}_{j_p m_p j_n m_n}^{JM}$ are Clebsh-Gordan coefficients.

The commutator $[A,A^{\dagger}]$ is replaced within pn-RQRPA by
its mean value in the QRPA vacuum
\begin{equation}
  \begin{split}
    [A,A^{\dagger}] &\to
    \langle 0^+_{\mathrm{RPA}} | [A(p n,J M),
    A(p' n',J M)] | 0^+_{\mathrm{RPA}} \rangle \\
    &= \delta_{pp'} \delta_{nn'}\left\{1-\frac{1}{\hat{j}_p}\langle
    0^+_{\mathrm{RPA}}  | [a^{\dagger}_{p} \tilde{a}_p]_{00} |
    0^+_{\mathrm{RPA}} \rangle - \frac{1}{\hat{j}_n}\langle
    0^+_{\mathrm{RPA}}  | [a^{\dagger}_{n} \tilde{a}_n]_{00} |
    0^+_{\mathrm{RPA}} \rangle\right\}\\
    &\equiv \delta_{pp'} \delta_{nn'} D_{pn,J^{\pi}},
    \label{eq:rqrpa-comut}
  \end{split}
\end{equation}
where $\hat{j}_p \equiv \sqrt{2j_p + 1}$ and
\begin{equation}
  [a^{\dagger}_{p} \tilde{a}_p]_{00} \equiv \sum\limits_{m_p}
  \mathrm{C}^{00}_{j_p m_p j_p -m_p} a^{\dagger}_{p m_p} a_{p -m_p}.
\end{equation}
Within the quasiboson approximation,
RPA vacuum $| 0^+_{\mathrm{RPA}} \rangle$
in Eq.~\eqref{eq:rqrpa-comut} is replaced by non-correlated BCS vacuum
$| 0^+_{\mathrm{BCS}}\rangle$ (i.e. $D_{pn,J^{\pi}} = 1$). Quasiboson
approximation violates Pauli exclusion principle.

From the Schr{\"o}dinger equation
\begin{equation}
  [H,Q_{JM^{\pi}}^{m \dagger}] | 0^+_{\mathrm{RPA}} \rangle
  = \Omega^m_{J^{\pi}} Q_{JM^{\pi}}^{m \dagger} | 0^+_{\mathrm{RPA}} \rangle,
\end{equation}
with the excitation energy $\Omega^m_{J^{\pi}}$, we obtain RQRPA
equation,
\begin{equation}
  \begin{pmatrix}
    \overline{\cal A} & \overline{\cal B}\\
    \overline{\cal B} & \overline{\cal A}
  \end{pmatrix}
  \begin{pmatrix}
    \overline{X}^m\\
    \overline{Y}^m
  \end{pmatrix}
  = \Omega^m_{J^{\pi}}
  \begin{pmatrix}
    \overline{X}^m\\
    -\overline{Y}^m
  \end{pmatrix}.
  \label{eq:rqrpa}
\end{equation}
Here matrices $\overline{\cal A}$, $\overline{\cal B}$ have the
form
\begin{equation}
  \begin{split}
    \overline{\cal A}^{J^{\pi}}_{pn,p'n'} &=
    (E_p + E_n)\delta_{pp'}\delta_{nn'} -
    2[G(pn,p'n';J)(u_p u_n u_{p'} u_{n'} +
    v_{p'} v_{n} v_{p'} v_{n'}) \\
    &+
    F(pn,p'n';J)(u_p v_n u_{p'} v_{n'} + v_{p} u_{n} v_{p'} u_{n'}) ]
    D^{1/2}_{pn,J^{\pi}} D^{1/2}_{p'n',J^{\pi}},
  \end{split}
\end{equation}
\begin{equation}
  \begin{split}
    \overline{\cal B}^{J^{\pi}}_{pn,p'n'} &= (E_p + E_n)
    2[G(pn,p'n';J)(u_p u_n v_{p'} v_{n'} +
    v_{p'} v_{n} u_{p'} u_{n'}) \\
    &- F(pn,p'n';J)(u_p v_n v_{p'} u_{n'} +
    v_{p} u_{n} u_{p'} v_{n'}) ]
    D^{1/2}_{pn,J^{\pi}} D^{1/2}_{p'n',J^{\pi}}
  \end{split}
\end{equation}
and amplitudes $\overline{X}^m_{(pn,J^{\pi})}$,
$\overline{X}^m_{(pn,J^{\pi})}$ are
\begin{equation}
  \overline{X}^m_{(pn,J^{\pi})} = D^{1/2}_{pn,J^{\pi}}
  X^m_{(pn,J^{\pi})},\quad
  \overline{Y}^m_{(pn,J^{\pi})} = D^{1/2}_{pn,J^{\pi}}
  Y^m_{(pn,J^{\pi})},
\end{equation}
where $F(pn,p'n';J)$ is the particle-hole interaction matrix
element. From the mapping procedure~(\ref{eq:rqrpa-comut}) we
obtain for the coefficients $D_{pn,J}$ the system of nonlinear
equations~\cite{sch96}
\begin{equation}
  D_{pn,J}=1-\frac{1}{\hat{j_p^2}}\sum\limits_{n'J'm} D_{pn',J'^{\pi}}
  |\overline{Y}^m_{(pn',J^{\pi})}|^2 -
  \frac{1}{\hat{j_n^2}}\sum\limits_{p'J'm} D_{p'n,J'^{\pi}}
  |\overline{Y}^m_{(p'n,J^{\pi})}|^2.
  \label{eq:D-map}
\end{equation}
The amplitudes $\overline{X}^m_{(pn,J^{\pi})}$,
$\overline{Y}^m_{(pn,J^{\pi})}$ and the excitation energies
$\Omega^m_{J^{\pi}}$ are obtained by iterating of the coupled
equations~(\ref{eq:D-map}) a (\ref{eq:rqrpa}).

The $(\mu^{-}, e^+)$ conversion nuclear matrix elements within
pn-RQRPA are transformed to the sum of the two-particle matrix
elements
\begin{equation}
  \begin{split}
    M^{\mathrm{type}} &=
    \sum\limits_{pnp'n' \atop{ J^{\pi}m_i m_f
    {\cal J}}} (-1)^{j_n + j_p'+ J + {\cal J}} (2{\cal J} + 1)
    \begin{Bmatrix}
      j_p  &j_n  &J\\
      j_n' &j_p' &{\cal J}
    \end{Bmatrix}\\
    &\times\langle p(1), p'(2); {\cal J} |
    f(r_{12}) \tau_{1}^+ \tau_{2}^+
    {\cal O}^{\mathrm{type}}_{12} f(r_{12}) | n(1), n'(2); {\cal J}
    \rangle \\
    &\times \langle 0^+_f || [\widetilde{c_{p'}^{\dagger},{\tilde c}_{n'}}]_J ||
    J^{\pi} m_f \rangle
    \langle J^{\pi} m_f| J^{\pi} m_i \rangle
    \langle J^{\pi} m_i || [c_{p}^{\dagger},\tilde{c}_{n}]_J || 0^+_i
    \rangle.
    \label{eq:M-qrpa}
  \end{split}
\end{equation}
Here $\{ \cdots \}$ is the Wigner $6j$ symbol, ${\cal
O}^{\mathrm{type}}_{12}$ is space- and spin-dependent part of the
matrix element. The single particle densities are defined as
\begin{equation}
\frac{
  \langle 0^+_f ||
  [c_{p}^{\dagger},\tilde{c}_{n}]_J || 0^+_i \rangle }
  {\sqrt{2J + 1}} = (u_p^{(i)} v_n^{(i)}
  \overline{X}_{(pn,J^{\pi})}^{m_i} +
  v_p^{(i)} u_n^{(i)} \overline{Y}_{(pn,J^{\pi})}^{m_i})
  \sqrt{D_{pn,J^{\pi}}^{(i)}},
%
\end{equation}
\begin{equation}
  \frac{ \langle 0^+_f ||
  [\widetilde{c_{p}^{\dagger},{\tilde c}_{n}}]_J || 0^+_i
  \rangle }{\sqrt{2J + 1}}
  = (v_p^{(f)} u_n^{(f)} \overline{X}_{(pn,J^{\pi})}^{m_f} +
  u_p^{(f)} v_n^{(f)} \overline{Y}_{(pn,J^{\pi})}^{m_f})
  \sqrt{D_{pn,J^{\pi}}^{(f)}},
\end{equation}
where the indices $(i)$ and  $(f)$ indicate that the excitations
are defined with the respect to the ground state of the initial
and final nucleus respectively. When these states are not the
same, the overlap factor
\begin{equation}
  \langle J^{\pi} m_f| J^{\pi} m_i \rangle \approx
  \sum\limits_{pn} (\overline{X}_{(pn,J^{\pi})}^{m_i}
  \overline{X}_{(pn,J^{\pi})}^{m_f} -
  \overline{Y}_{(pn,J^{\pi})}^{m_i} \overline{Y}_{(pn,J^{\pi})}^{m_f})
  (u^{(i)}_n u^{(f)}_n + v^{(i)}_n v^{(f)}_n).
\end{equation}
must be introduced~\cite{sim98}. Repulsion between the nucleons at
short distances is described by the short-range correlation factor
$f(r_{12})$ of the form
\begin{equation}
  \label{eq:f12}
  f(r_{12}) = 1- \mathrm{e}^{- \alpha r^2_{12}}(1-b r^2_{12}),
\end{equation}
where $\alpha = 1.1~\mathrm{fm}^2$ a $b = 0.68~\mathrm{fm}^2$ \cite{jastr}.
Particle-particle and particle-hole channels of the nuclear Hamiltonian are
renormalized by the parameters $g_{pp}$ and $g_{ph}$:
\begin{align}
  F(pn,p'n';J) &\to g_{ph} F(pn,p'n';J),\\
  G(pn,p'n';J) &\to g_{pp} G(pn,p'n';J).
\end{align}


\end{document}